\documentclass[11pt,aps,prd,superscriptaddress,amsmath,amssymb,nofootinbib]{revtex4-2}
\usepackage[T1]{fontenc}
\usepackage[utf8]{inputenc}
\usepackage[colorlinks=true, allcolors=blue]{hyperref}
\usepackage{graphicx}
\usepackage{xcolor}

\definecolor{davi}{RGB}{0, 0 , 200}

\begin{document}


\title{Primordial perturbations and inflation in a holography inspired Gauss-Bonnet cosmology}
\author{Nicolas R.\ Bertini}\email{nicolas.bertini@cosmo-ufes.org}
\affiliation{Núcleo de Astrofísica e Cosmologia \& Departamento de F\'isica, Universidade 
Federal do Esp\'irito Santo. Av.\ Fernando Ferrari 514, 29075-910, Vit\'oria, ES, Brazil}
\author{Neven Bili\'c}\email{bilic@irb.hr}
\affiliation{Núcleo de Astrofísica e Cosmologia \& Departamento de F\'isica, Universidade 
Federal do Esp\'irito Santo. Av.\ Fernando Ferrari 514, 29075-910, Vit\'oria, ES, Brazil}
\affiliation{Division of Theoretical Physics, Rudjer Bo\v skovi\'c Institute, 10002 Zagreb, Croatia}
\author{Davi C.\  Rodrigues}\email{davi.rodrigues@cosmo-ufes.org}
\affiliation{Núcleo de Astrofísica e Cosmologia \& Departamento de F\'isica, Universidade 
Federal do Esp\'irito Santo. Av.\ Fernando Ferrari 514, 29075-910, Vit\'oria, ES, Brazil}

\begin{abstract}

We consider an action for gravity that, in addition to the Einstein-Hilbert term,
contains a function of the Ricci scalar and the Gauss-Bonnet invariant. The specific form of the function 
considered is motivated by holographic cosmology.
At background level the field equations imply modified Friedmann equations of the same form as
those in the holographic cosmology.
We calculate the cosmological perturbations and derive the corresponding power spectra assuming a general $k$-inflation. 
We find that the resulting power spectra differ substantially  from those
obtained in both  holographic and standard cosmology. The estimated spectral index and tensor-to-scalar ratio are confronted with the Planck results.
\end{abstract}

\maketitle

\section{Introduction}
A modified Gauss-Bonnet (MGB) gravity \cite{cognola,nojiri,nojiri2,nojiri3,elizalde2}
is a class of modified gravity models in which
the gravitational action is a general function of two variables: the Ricci scalar $R$ and  the Gauss-Bonnet invariant
\begin{equation}
\mathcal{G}=R^2 -4R^{\mu\nu}R_{\mu\nu}+R^{\mu\nu\rho\sigma}R_{\mu\nu\rho\sigma}.
 \label{eq1001}
\end{equation}
The functional dependence on $R$ and $\mathcal{G}$ can be further constrained  by physical requirements.
In a cosmological context  it is natural to
require that the second Friedmann equation is linear
in $\dot{H}$.
Then, in addition to the Einstein-Hilbert term the gravitational Lagrangian can contain
a function of $R$ and $\mathcal{G}$ of the form
$f(J)$ depending only on one invariant \cite{gao}
 \begin{equation}
J= \frac{1}{\sqrt{12}}
\left(-R+\sqrt{R^2-6\mathcal{G}}
\right)^{1/2}
 \label{eq1003}
\end{equation}
normalized so that $J^2=H^2$ for a spatially flat cosmology at background level. 
The Friedmann equations in this case become very simple:  the left-hand side of the first Friedmann equation, in addition to the usual $H^2$ term contains
a general function of $H$.


In this paper we study 
in particular a MGB gravity with $f(J) \propto J^4$.
 In the following we will refer to this theory as the MGB model.
 In this model
 one obtains the cosmology equivalent to that on the holographic braneworld
\cite{apostolopoulos,bilic1,bilic3,bertini} at background level.
This equivalence between the two models poses a natural question if the equivalence goes beyond background
cosmology. In particular, it would be of considerable interest to check if the two models
produce similar spectra of the primordial cosmological perturbations.

At this point it is worth mentioning a few related works
in the context of inflation. 
In Ref.\ \cite{delcampo}  inflationary models were studied with  arbitrary functions of $H$ added to the 
left-hand side of the  first Friedmann equation. 
Chackraborty et al \cite{Chakraborty:2018scm} have studied inflation in a model  with the Gauss-Bonnet term coupled to a scalar field.
Basilacos et al \cite{Basilakos:2020qmu} have shown that a modification of the Friedmann equation with a quartic term $\propto H^4$ is obtained 
in a string theory inspired model with a Kalb-Ramond term in the Lagrangian.
In spite of some similarities, these examples are not equivalent to the model considered here.

	Inflation in the modified gravity models in  
	which the gravitational Lagrangian is a function  $F(R,\mathcal{G})$ 
	have been studied in Refs.\ \cite{delaurentis,odintsov4}.
	De Laurentis et al \cite{delaurentis} have studied inflation in a kind of Starobinski extended models of the type
	$F(R,\mathcal{G})=- R+a R^2+ b\mathcal{G}$ where $a$  and $b$ are constants.
	Odintsov et al \cite{odintsov4} have studied inflation in models of the type
	$F(R,\mathcal{G})=- R+a \mathcal{G}^\alpha$ where $\alpha\neq 1$.
	In both Refs. \cite{delaurentis,odintsov4} the action does not contain matter fields and inflation is driven  
	solely by the geometry.  In our approach, in contrast, inflation is driven by a scalar field coupled to the modified
	gravity with Lagrangian of the type $F(R,\mathcal{G})= -R+(\ell^2/2)J^4$
	where the invariant $J$ is given by (\ref{eq1003}) and
	$\ell$ is a constant of dimension of length.

In a recent paper \cite{bertini} we have presented the calculations of the cosmological perturbations for a $k$-essence
field theory in the holographic braneworld in the context of inflation.
We have demonstrated that the perturbations produce
the power spectra as in the standard $k$-inflation in general relativity.
Here we calculate the perturbations in the
MGB model with background equations identical to the holographic cosmology and
find  a substantial departure from the general-relativistic results.

	As a side issue, it is important to address the ghost instability problem in MGB models
	which is somewhat controversial. 
	It was argued that the modified gravity models in  
	which
	the gravitational action is a general function  $F(R,\mathcal{G})$ 	 
	are ghost free \cite{comelli,navarro}.
	However, in a recent paper \cite{felice} it was demonstrated that, with a few exceptions, there is an instability in the scalar sector of   $F(R,\mathcal{G})$ models. We present a brief review of these issues in Appendix \ref{ghost}, where we also point out why the analysis of Ref.\ \cite{felice} does not apply to the model considered here.

The remainder of the paper is organized as follows.
In Sec.\ \ref{field} we introduce the MGB model and  using the scalar-tensor representation formalism we derive the background field equations
from which we derive the corresponding Friedmann equations.
In Sec.\ \ref{perturbations} we derive the spectra of the cosmological perturbations for the MGB model with  $k$-essence. In Sec.\ \ref{index} we calculate the power spectra and spectral indices. 
Concluding remarks are given in Sec.\ \ref{conclude}.
In Appendix \ref{justify} we justify some approximations made in Sec.\ \ref{perturbations}.
In appendices  \ref{app-scalar} and \ref{app-tensor} we present details of the calculations of the scalar and tensor perturbations, respectively.
Appendix \ref{ghost} is devoted to  ghost issues in general $F(R, \mathcal{G})$ theories and to the particular case  considered here.

\section{Field equations in the MGB model}
 \label{field}

\subsection{The action}

Consider the MGB action of the the form
\begin{equation}
S_{\rm MGB}=\int d^4x\sqrt{-g}\left[
\frac{1}{16\pi G_{\rm N}}(-R+f(J))+\mathcal{L}
\right],
 \label{eq1002}
\end{equation}
where  $G_{\rm N}$ is the Newtonian constant and
$f$ is a smooth function of the invariant $J$ defined by (\ref{eq1003}).
 We will assume that the value of $G_{\rm N}$  is provided by the measurements in the the solar system since the modifications of gravity should be relevant only for short distances. Matter is represented by
a Lagrangian $\mathcal{L}$ as a general function of the scalar field $\theta$ and kinetic term
\begin{equation}
X\equiv g^{\mu\nu}\theta_{,\mu}\theta_{,\nu} .
\label{eq0201}
\end{equation}
This type of scalar field theories,  dubbed $k$-essence \cite{armendariz,armendariz2}, is very general
 and includes the
canonical scalar field theory as a particular case. A  $k$-essence  is dynamically equivalent to a generally non-isentropic and non-barotropic potential fluid flow, whereas
a purely kinetic $k$-essence is equivalent to a barotropic potential flow \cite{arroja,Piattella:2013wpa}.

 For a general Friedmann-Lema\^itre-Robertson-Walker (FLRW) metric with line element
\begin{equation}
ds^2=g^{(0)}_{\mu\nu}dx^\mu dx^\nu =dt^2 -a^2(t) \left[
d\chi^2+\frac{\sin^2(\sqrt{\kappa}\chi)}{\kappa}(d\vartheta^2+\sin^2 \vartheta d\varphi^2)
\right],
 \label{eq3201}
\end{equation}
one finds
\begin{equation}
 J^2=H^2+\frac{\kappa}{a^2} ,
 \label{eq6006}
\end{equation}
where $\kappa=$ 1, -1, or 0 for
closed, open hyperbolic, or open flat space, respectively.
Then applying the  Euler-Lagrange formalism (for some efficient methods see Refs.\ \cite{gao, Casalino:2019tho}), we find a modified first Friedmann equation in the form
\begin{equation}
 J^2+\frac16 \left( f(J) - H\frac{\partial f}{\partial H}\right)
 =\frac{8\pi G_{\rm N}}{3}\rho .
 \label{eq1006}
\end{equation}
Hence, the left hand side is a function of $a$ and $H$ only and the second Friedmann equation
will be linear in $\dot{H}$.  The above equation extends the $\kappa=0$ result of Ref.\ \cite{gao} to arbitrary $\kappa$ values (see also Ref.\ \cite{Casalino:2019tho}). In the next section, the Friedmann equations are derived directly from the field equations with a specific function $f(J)$.

An interesting particular case is obtained for
\begin{equation}
f(J)=\frac12 \ell^2 J^4,
 \label{eq1008}
\end{equation}
where  $\ell$ is a coupling constant of dimension of length.
In this case the Friedman equation (\ref{eq1006}) takes the form obtained in the
spatially flat holographic
cosmology \cite{apostolopoulos,bilic1,bertini} if we identify the constant $\ell$
with the AdS$_5$ curvature radius. In the following, we will study the action
(\ref{eq1002}) with (\ref{eq1008}), i.e.,
\begin{equation}
	S= \frac{1}{16\pi G_{\rm N}}\int d^{4}x\sqrt{-g}F(R,\mathcal{G})+\int d^{4}x\sqrt{-g}\mathcal{L} ,
\label{eq:2}
\end{equation}
where
\begin{equation}
F(R,\mathcal{G}) = -R  + \frac{\ell^{2}}{288} \Big(-R +\sqrt{R^{2}-6 \mathcal{G} } \Big)^{2}.
\label{eq:3}
\end{equation}

\subsection{Scalar-tensor representation}
\label{scalar-tensor}
 It is well known that $f(R)$ gravity can be  described by a dual scalar-tensor action with a single scalar (for a review see, e.g., \cite{Capozziello:2010zz}). More general extended gravity theories, as is the action \eqref{eq:2}, may need additional scalars \cite{Wands:1993uu, rodrigues} .  Here we follow the approach of \cite{rodrigues} to find the dual action.
For a general gravitational action with an arbitrary dependence on some metric invariants $X^{a}$
(e.g., $R$ and $\mathcal{G}$)
\begin{equation}
S_1= \int d^{4}x\sqrt{-g}F(X_{a}),
\end{equation}
we can form a dual action
by making use of a Legendre transformation
\begin{equation}
S_2= \int d^{4}x\sqrt{-g}(\psi_{a}X^{a}-V(\psi^{a})),
\end{equation}
where $\psi_{a}$ are scalar fields that satisfy the following relations
\begin{equation}
\psi_{a} = \frac{\partial F}{\partial X^{a}}; \qquad X^{a} = \frac{\partial V}{\partial \psi_{a}}.
\label{eq5400}
\end{equation}
For the action \eqref{eq:2} with \eqref{eq:3} the dual version is
\begin{equation}
S_{D} = \frac{1}{16\pi G_{\rm N}}\int d^{4}x\sqrt{-g}\Bigg( \psi_{1}R+\psi_{2}\mathcal{G} -  V(\psi_{1},\psi_{2}) \Bigg)
+\int d^{4}x\sqrt{-g}\mathcal{L} ,
\end{equation}
where
\begin{equation}
	\psi_{1} =\frac{\partial F (R,\mathcal{G}) }{\partial R} =
-1-\frac{\ell^2  (-R+\sqrt{R^2-6 \mathcal{G}})^2}{144\sqrt{R^2-6 \mathcal{G}}},
\label{eq5301}
\end{equation}
\begin{equation}
\psi_{2} = \frac{\partial F (R,\mathcal{G})}{\partial \mathcal{G}}  =
\frac{\ell^2}{48}  \left(-1+\frac{ R}{\sqrt{R^2-6 \mathcal{G}}}\right).
\label{eq5302}
\end{equation}
Then, the second set of equations in  (\ref{eq5400}) reads
\begin{equation}
R= \frac{\partial V}{\partial \psi_1}, \quad \mathcal{G}= \frac{\partial V}{\partial \psi_2},
\label{eq5401}
\end{equation}
and by integrating these 
we obtain
\begin{equation}
V(\psi_{1},\psi_{2}) = -\frac{3 \left(\psi _1+1\right)^2 \left(\psi _2+ \ell^{2}/48 \right)}{2 \psi _2^2}.
\end{equation}

The variation of $S_{D}$ with respect to the metric leads to modified Einstein's equations
\begin{eqnarray}
-\psi_{1}G_{\mu\nu}-\frac{1}{2}g_{\mu\nu} V(\psi_{1},\psi_{2})  -g_{\mu\nu}\Box\psi_{1}+\nabla_{\mu}\nabla_{\nu}\psi_{1}
+4(\Box\psi_{2})G_{\mu\nu}
\nonumber\\
+2(\nabla_{\mu}\nabla_{\nu}\psi_{2})R- 4(\nabla_{\rho}\nabla_{\mu}\psi_{2})R_{\nu}^{\;\rho}
-4(\nabla_{\rho}\nabla_{\nu}\psi_{2})R_{\mu}^{\;\rho}
\nonumber\\
 + 4g_{\mu\nu}(\nabla_{\rho}\nabla_{\sigma}\psi_{2})R^{\sigma\rho}
-4(\nabla^{\rho}\nabla^{\sigma}\psi_{2})R_{\mu\rho\nu\sigma}=8\pi G_{\rm N}T_{\mu\nu},
\label{eq:mod}
\end{eqnarray}
where the energy momentum tensor $T_{\mu\nu}$ is associated with the matter Lagrangian $\mathcal{L}$. 
Note that the variation of $\cal G$ in the above expression  is in agreement with Ref.\ \cite{Nojiri:2005vv}. 
One can easily check that the variation with respect to $\psi_1$ and $\psi_2$ yields a pair of equations equivalent to
(\ref{eq5301}) and (\ref{eq5302}).

\subsection{Background equations}
\label{background}
Now we specify the background metric to the FLRW form (\ref{eq3201})
and we assume
\begin{equation}
{T}^{\mu}_{\nu}=\mbox{diag}(\rho,- p,-p,-p) .
\label{eq3010}
\end{equation}
Then, using the modified Einstein equations (\ref{eq:mod})  we obtain the following
modified Friedmann equations
\begin{equation}
3\left( H^{2} + \frac{\kappa}{a^{2}} \right) - \frac{\ell^{2}}{4}\left( H^{2}+\frac{\kappa}{a^{2}}\right)
\left( 3H^{2} - \frac{\kappa}{a^{2}} \right) = 8\pi G_{\rm N} \rho ,
 \label{eq3110}
\end{equation}
\begin{equation}
\dot{H}\left[ 1-\frac{\ell^{2}}{6}\left( \frac{\kappa}{a^{2}}+ 3H^{2} \right) \right] - \frac{\kappa}{a^{2}}
\left[1 - \frac{\ell^{2}}{6}\left( H^{2} - \frac{\kappa}{a^{2}} \right) \right] = - 4\pi G_{\rm N} (p + \rho)  .
 \label{eq4111}
\end{equation}
Of course, Eq.\ (\ref{eq3110}) agrees with Eq.\ (\ref{eq1006}) for $f(J)$ given by (\ref{eq1008}).
It is easy to show that Eqs.\ (\ref{eq3110}) and (\ref{eq4111}) imply
\begin{equation}
\dot{\rho}+3H(p+\rho)=0,
 \label{eq3109}
\end{equation}
which also follows from energy-momentum conservation
\begin{equation}
{T^{\mu\nu}}_{;\nu}=0.
\end{equation}

In the following we adopt the usual assumption that the early universe is spatially flat. 
Then, Eqs.\ (\ref{eq3110}) and (\ref{eq4111})
with $\kappa=0$ reduce to
 \begin{equation}
H^2-\frac{\ell^2}{4}H^4=\frac{8\pi G_{\rm N}}{3} \rho ,
\label{eq0002}
\end{equation}
\begin{equation}
\dot{H}(1-\ell^2 H^2/2)=-4\pi G_{\rm N}(p+ \rho) ,
\label{eq0102}
\end{equation}
precisely as in the spatially flat holographic cosmology
\cite{apostolopoulos,bilic1,bertini}.

The pressure $p$ and energy  density  $\rho$
are derived from
$\mathcal{L}$
using the usual prescription
  \begin{equation}
p=\mathcal{L}, \quad \rho= 2X{\mathcal{L}}_{,X}-{\mathcal{L}},
\label{eq0008}
\end{equation}
where the kinetic term $X$ is defined in (\ref{eq0201})
and the subscript $_{,X}$ denotes a partial derivative with respect to $X$.
The energy-momentum tensor is then given by
\begin{equation}
 {T}_{\mu\nu}=( {p}+ {\rho}) u_\mu u_\nu - {p}g_{\mu\nu} ,
\label{eq0005}
\end{equation}
where
\begin{equation}
u_\mu =\frac{\theta_{,\mu}}{\sqrt{X}}.
\label{eq0010}
\end{equation}

The MGB cosmology has interesting properties. Solving the first
Friedmann equation (\ref{eq0002}) as a quadratic equation for $H^2$ we find
\begin{equation}
H^2=\frac{2}{\ell}\left(1\pm\sqrt{1-\frac{8\pi G_{\rm N}\ell^2}{3}\rho}\right).
 \label{eq125}
\end{equation}
Now, by demanding that Eq.\ (\ref{eq125})
reduces to the standard Friedmann
equation in the low density limit, i.e., in the limit when
$G_{\rm N}\ell^2\rho \ll 1$, we are led to keep
only the ($-$) sign solution in (\ref{eq125})
and discard the ($+$) sign solution as unphysical.
Then, it follows
that the physical range of the Hubble expansion rate is between zero and the
maximal value $H_{\rm max}= \sqrt2/\ell$ corresponding to the maximal energy density
$\rho_{\rm max}= 3/(8\pi G_{\rm N}\ell^2)$ \cite{bilic1,delcampo}.
Assuming no violation of the weak energy condition
$p+\rho\geq 0$,
 the expansion rate will, according to (\ref{eq0102}), be a monotonously decreasing function of time.
 
 Our ambition here is by no means an attempt to explain the very beginning of the  universe. Nevertheless, it is worth noting that if the evolution starts from $t=0$ with an initial $H_{\rm i} \leq H_{\rm max}$
 the initial energy density and cosmological expansion scale will be both finite.
Hence, as already noted by Gao \cite{gao}, in the  modified cosmology described
by the Friedmann equations (\ref{eq0002}) and (\ref{eq0102}),
the Big Bang singularity is avoided.

 The expansion of the early universe is conveniently described
using the so called slow-roll parameters.
We use the following recursive definition of the slow-roll parameters
\cite{schwarz,steer}
\begin{equation}
\varepsilon_{i+1}=\frac{\dot{\varepsilon}_i}{H\varepsilon_i} ,
 \label{eq3214}
\end{equation}
starting with
\begin{equation}
\varepsilon_1= -\frac{\dot{H}}{H^2}.
 \label{eq3114}
\end{equation}
The beginning of inflation is characterized by the slow-roll regime
with slow-roll parameters satisfying $\varepsilon_i\ll 1$.

\subsection{Speed of sound}
\label{sound}
The adiabatic speed of sound is given by
\begin{equation}
c_{\rm s}^2\equiv \left.\frac{\partial p}{\partial\rho}\right|_\theta =\frac{p_{,X}}{\rho_{,X}}
=\frac{p_{,X}}{p_{,X}+2Xp_{,XX}}
=\frac{p+\rho}{2X \rho_{,X}}.
\label{eq0018}
\end{equation}
In the slow-roll regime, the sound speed deviates slightly from unity and may be expressed in terms of
the slow-roll parameters $\varepsilon_{i}$.
First, by making use of the definition (\ref{eq3114}) and modified Friedman equations (\ref{eq0002}), (\ref{eq0102})
with (\ref{eq0008}),
we can express
the variable $X$ in the slow-roll regime as
\begin{equation}
X=- \frac{2p(2-h^2)}{3p_{,X}(4-h^2)}\varepsilon_1+ \mathcal{O}(\varepsilon_i^2),
\label{eq5107}
\end{equation}
where we have abbreviated
\begin{equation}
h\equiv \ell H.
 \label{eq2015}
\end{equation}
Then from (\ref{eq0018}) we find
\begin{equation}
c_{\rm s}^2=1+ \frac{4(2-h^2)}{3(4-h^2)}\frac{p p_{,XX}}{p_{,X}^2}\varepsilon_1+ \mathcal{O}(\varepsilon_i^2).
\label{eq5108}
\end{equation}
For example, in the tachyon model with
Lagrangian $\mathcal{L}=-V\sqrt{1-X}$ one finds \cite{bilic3}
\begin{equation}
c_{\rm s}^2=1-\frac{4(2-h^2)}{3(4-h^2)}\varepsilon_1+ \mathcal{O}(\varepsilon_i^2).
\label{eq5109}
\end{equation}

\section{Perturbations in MGB gravity}
\label{perturbations}

Here we derive the spectra of the cosmological perturbations for the
MGB cosmology
 with matter represented by a general $k$-essence.
 We shall closely follow J.\ Garriga and V.\ F.\ Mukhanov  \cite{garriga}
 and adjust their formalism to account for the modification of the Einstein equations.

 \subsection{Scalar perturbations}
 \label{scalar-perturbations}
Assuming a spatially flat background with line element (\ref{eq3201}) with $\kappa=0$, we introduce
the perturbed line element in the Newtonian gauge
\begin{equation}
 ds^2=(1+2\Psi) dt^2-(1-2\Phi)a^2(t)(dr^2+r^2 d\Omega^2) .
 \label{eq0013}
\end{equation}
 Inserting the above metric components in the field equations (\ref{eq:mod}) we obtain
a set of equations for $\Phi$ and $\Psi$
derived in appendix \ref{app-scalar}.
The relevant equations are (\ref{eq:02}), (\ref{eq:03}),  and the off-diagonal part of \eqref{eq:4}.
Owing to  $\delta T^{i}_{j}=\delta^{i}_{j}\delta p$
the off-diagonal part of Eq.\ \eqref{eq:4} in momentum space can be written as
\begin{align}
	\left[ 1 - \frac{h^{2}}{3}\left( \frac{H^{2}}{2\dot{H}} - \frac{\dot{H}}{H^{2}} + \frac{\ddot{H}}{H\dot{H}} - \frac{\ddot{H}^{2}}{\dot{H}^{3}} + \frac{\overset{...}{H}}{2\dot{H}^{2}}  \right) \right]\Phi
	\nonumber\\
	- \left[  1 - \frac{h^{2}}{6}\left(\frac{H\ddot{H}}{\dot{H}^{2}} - \frac{H^{2}}{\dot{H}} - 2 \right) \right]\Psi
	\nonumber\\
	+\frac{h^{2}}{9H^{2}}\frac{k^{2}}{a^{2}}\Phi + \frac{h^{2}}{18 \dot{H}}\frac{k^{2}}{a^{2}}\Psi
	\nonumber\\
	-\frac{h^{2}}{3}\left( \frac{H}{\dot{H}} - \frac{1}{H} \right)\dot{\Phi} -\frac{h^{2}H}{6\dot{H}}\dot{\Psi} = 0,	
\label{eq:off-diag}
\end{align}
where $h=\ell H$.
Hence, the slip parameter  defined in momentum space as 
\begin{equation}
	\eta\equiv \frac{\Phi}{\Psi}
	\label{eq:slip}
\end{equation}
is in general a function of  $k$ and $t$
and can be calculated numerically for a specific inflation model. 
However, making use of the slow-roll parameters will prove helpful to develop a model independent estimate on $\eta$ at horizon crossing (i.e., $k=aH$).
%
First,  with the help of (\ref{eq3214}) and (\ref{eq3114}),
Eq.
(\ref{eq:off-diag}) with $k=aH$  becomes
\begin{align}
	\left(1+ \frac{3+\varepsilon_1 (2-3 \varepsilon_2)-3 \varepsilon_2^2+3 \varepsilon_2 \varepsilon_3}{18 \varepsilon_1} h^2 \right)\Phi - 
	\left(1- \frac{2-3 \varepsilon_2 }{18 \varepsilon_1}h^2 \right)\Psi 
	 \nonumber \\
	+ \frac{1+\varepsilon_1}{3 \varepsilon_1 H } h^2 \dot{\Phi }+\frac{1}{6 \varepsilon_1 H } h ^2 \dot{\Psi }=0.
	\label{eq1000}	
\end{align}
In the slow-roll regime,  it is reasonable to assume 
\begin{equation}
	\dot{f} \simeq \mathcal{O}(\varepsilon_1)H  f ,
	\label{eq5300}
\end{equation}
where $f$ stands for an arbitrary smooth and slow varying function of time.
 For example, $\dot{H}=-\varepsilon_1 H^2$, $\dot{\varepsilon}_1=\varepsilon_2 H\varepsilon_1$, etc.  
 In view of the above relation, we introduce arbitrary parameters 
 $\varepsilon_\Phi$ and $\varepsilon_\Psi$ of order $ \mathcal{O}(\varepsilon_1)$ and write 
\begin{equation}
 \dot \Phi = \varepsilon_\Phi H \Phi, \quad \dot \Psi = \varepsilon_\Psi H \Psi.
 \label{eq6301}
\end{equation} 
Then we find
\begin{equation}
	\eta = 
	\frac{18 \varepsilon_1 - (2 - 3 \varepsilon_2 + 3 \varepsilon_\Psi) h^2}
	{18 \varepsilon_1 - [(-2 + 3 \varepsilon_2 - 6 \varepsilon_\Phi)\varepsilon_1 + 3(-1 + 
		\varepsilon_2^2 - \varepsilon_2 \varepsilon_3 - 2 \varepsilon_\Phi ) ] h^2} \, .
\label{eq900}
\end{equation}
The above expression is  exact on the
	proviso that $\Phi$ and $\Psi$ satisfy (\ref{eq6301}). Note that $\eta \rightarrow 1$
as $h\rightarrow 0$, as expected.
Now we make an approximation by
taking  all epsilons to be nearly equal, i.e., $\varepsilon_1\simeq\varepsilon_2\simeq\varepsilon_3\simeq\varepsilon_\Phi\simeq\varepsilon_\Psi\simeq\varepsilon$. Then  we obtain
\begin{equation} \label{etaEqual}
	\eta \simeq \frac{18 \varepsilon -2 h^2}{18 \varepsilon +\left( 3 + 8 \varepsilon  + 3 \varepsilon ^2\right) h^2} \, .
\end{equation}
The above function $\eta=\eta(\varepsilon,h)$ has a single minimum and a single maximum which yields the lower and upper bounds on $\eta$ as
\begin{equation}
	- \frac 23 < \eta < 1  .
\end{equation} 
The  maximum is found for $h \to 0$, while the minimum  for $\varepsilon \to 0$. If we require $\eta>0$,  Eq.\ \eqref{etaEqual} implies 
\begin{equation}
	\varepsilon > \frac{h^2}{9} .
\end{equation}
For sufficiently small $\varepsilon$ we have
\begin{equation}
	\eta \simeq -\frac {2}{3} + \frac {2}{9}(45 + 8 h^2) \frac{\varepsilon}{h^2} + O\left(\varepsilon^2 \right) .
\end{equation}

In the intermediate slow-roll regime ($\varepsilon_i\sim h^2$) one can calculate $\eta$ numerically
for a specific model of $k$-essence. However, it is possible to obtain a rough
model independent estimate  in the intermediate slow-roll regime assuming as above $\varepsilon_1\simeq\varepsilon_2\simeq\varepsilon_3\simeq\varepsilon_\Phi\simeq\varepsilon_\Psi\simeq\varepsilon$.
Let $\varepsilon_h$ denote the value of $\varepsilon$ close to $h^2$. Then
\begin{equation}
	\eta|_{\varepsilon \sim h^2} \simeq \frac{16}{21 + 8 \varepsilon_h + 3 \varepsilon_h^2} ,
\end{equation}
and hence,  we have  $\eta|_{\varepsilon \sim h^2}>0$. Moreover, assuming that inflation ends when $\varepsilon \sim 1$, then  $\eta|_{\varepsilon \sim h^2} > 1/2$.

The above estimates rely on the assumption that all of the epsilons are nearly equal. Nonetheless, it provides a simple and illustrative analytical description.
In the following we will consider yet another  approximation:  we will adopt the simplification that
during inflation $\eta$ can be taken to be a constant between 0 and 1.
%
\subsection{Scalar power spectrum}
\label{scalar}
Using the definition (\ref{eq:slip}) we can express the remaining perturbation equations
in terms of $\Phi$ and $\eta$.
The  perturbations of  the stress tensor components $\delta {T}^\mu_\nu$ are induced by the perturbations of
the scalar field $\theta(t,x)=\theta(t)+\delta\theta(t,x)$ and the perturbation of the metric.
Using the energy conservation (\ref{eq3109}) and the definition (\ref{eq0201}) of $X$
one finds
\begin{equation}
	\delta {T}^0_0=\frac{{p}+{\rho}}{{c}_{\rm s}^2}
	\left[\left(\frac{\delta\theta}{\dot{\theta}}\right)^{\mbox{.}}-\Psi\right]
	-3H({p}+{\rho}) \frac{\delta\theta}{\dot{\theta}},
	\label{eq0041}
\end{equation}
\begin{equation}
	\delta {T}^0_i=({p}+{\rho})\left(\frac{\delta\theta}{\dot{\theta}}\right)_{,i} ,
	\label{eq0043}
\end{equation}
where the adiabatic sound speed ${c}_{\rm s}$ is defined by (\ref{eq0018}).
Using (\ref{eq0043}) equation (\ref{eq:03}) becomes
\begin{align}
	\left(2-h^{2} \right)(\dot{\Phi}+ H\Psi)  - \frac{h^{2}}{9}\left( \frac{H}{\dot{H}} + \frac{\ddot{H}}{\dot{H}^{2}} - \frac{4}{H}\right)\frac{\nabla^{2}\Phi}{a^{2}} + \frac{h^{2}}{9\dot{H}} \frac{\nabla^{2}(\dot{\Phi}+H\Psi)}{a^{2}} = 8\pi G_{N}
	 ({p}+{\rho})\frac{\delta\theta}{\dot{\theta}}.
	 \label{eq:003}
\end{align}
Multiplying this by $3H$ and adding to (\ref{eq:02}) with (\ref{eq0041}) we obtain
\begin{align}
 2\left(  1-\frac{h^{2}}{6} \right)\frac{\nabla^{2}\Phi}{a^{2}}- \frac{ H h^{2}}{3 \dot{H}}\frac{\nabla^{2}(\dot{\Phi} + H\Psi)}{a^{2}}
 -\frac{h^{2}}{3}\left( \frac{H}{\dot{H}} + \frac{\ddot{H}}{\dot{H}^{2}} - \frac{4}{H}\right)\frac{\nabla^{2}\Phi}{a^{2}}
	\nonumber\\
-\frac{h^{2}}{9\dot{H}}\frac{k^{2}\nabla^{2}\Phi}{a^{4}} = 8\pi G_{N}
\frac{{p}+{\rho}}{{c}_{\rm s}^2}
\left[\left(\frac{\delta\theta}{\dot{\theta}}\right)^{\mbox{.}}-\Psi\right].	
\label{eq:0002}
\end{align} 
Here, in the last term on the left-hand side we have made a replacement
$\nabla^{2}\rightarrow -k^2$.
Next, employing  the slow-roll condition (\ref{eq5300})
we neglect $\dot{\Phi}$ in the second term, use the horizon crossing relation $k=aH$,  replace $\Psi$ by $\Phi/\eta$
and approximate $\eta$ by a constant, as discussed at the end of Sec.\ \ref{scalar-perturbations}.
Then,  by making use of the definitions  (\ref{eq3214}) and (\ref{eq3114}), from (\ref{eq:0002}) we obtain
\begin{align}
\gamma\frac{\nabla^2\Phi}{a^{2}} = 4\pi G_{N}
	\frac{{p}+{\rho}}{{c}_{\rm s}^2}
\left[\left(\frac{\delta\theta}{\dot{\theta}}\right)^{\mbox{.}}-\frac{\Phi}{\eta}\right],	
	\label{eq:0003}
\end{align} 
where
\begin{equation}
	\gamma=1+\frac{h^2}{6}  \left( 1+  \frac{\varepsilon_2}{\varepsilon_1}\right) + \left( \frac29 + \frac{1}{6\eta}\right)\frac{h^2}{\varepsilon_1}.
\label{eq0004}
\end{equation}
This is our first basic equation. The second equation is obtained from (\ref{eq:03}) in  which we replace 
$\nabla^2\rightarrow -a^2H^2$. Then we find 
\begin{equation}
(\alpha\dot{\Phi}+\beta H\Phi)_{,i}=4\pi G_{N}\delta T^{0}_{i},
\label{eq2006} 
\end{equation}
where 
\begin{equation}
\alpha=1-\frac{h^2}{6}+\frac{h^2}{18}\frac{1}{\varepsilon_1} ,
\end{equation}
\begin{equation}
\beta = \frac{1}{\eta}-h^2 \left(\frac19+\frac{1}{2\eta}\right)
-\frac{h^2}{18}
\left( 1-\frac{1}{\eta}\right)\frac{1}{\varepsilon_1}
-\frac{h^2}{18}\frac{\varepsilon_2}{\varepsilon_1} .
\end{equation}
Next, by noting that $\beta/\alpha = \mathcal{O}(1)$ and employing the slow-roll condition (\ref{eq5300}) we can neglect the first term in brackets on the left-hand side of
(\ref{eq2006}). Furthermore we use
\begin{equation}
	(a\Phi)^{\mbox{.}}=aH\Phi+a\dot{\Phi}\simeq aH\Phi
\end{equation}
and finally obtain
\begin{equation}
\beta (a\Phi)^{\mbox{.}} \simeq 4\pi G_{N} a
({p}+{\rho})\frac{\delta\theta}{\dot{\theta}} .
	\label{eq0006} 
\end{equation}

Now, we can proceed in a way similar to  Ref.\ \cite{garriga} (for more details see also \cite{bertini} and the appendix of \cite{bilic3}).
Introducing  
\begin{equation}
	\tilde{c}^2=\frac{\gamma}{\beta}c_{\rm s}^2 ,
\end{equation}
equations (\ref{eq:0003}) and (\ref{eq0006}) can be put in the form 
\begin{equation}
  \left(\frac{\delta\theta}{\dot{\theta}}\right)^{\mbox{.}}
=\frac{\Phi}{\eta} + \frac{\tilde{c}^2}{4\pi G_{\rm N}a^2(p+\rho)}\beta\nabla^2 \Phi ,
 \label{eq5012}
\end{equation}
\begin{equation}
(a\Phi)^{\mbox{.}}=4\pi G_{\rm N} a(p+\rho)\frac{1}{\beta}\frac{\delta\theta}{\dot{\theta}}.
 \label{eq5013}
\end{equation}
As shown in Appendix \ref{justify}, we can neglect the first term on the right-hand side of equation (\ref{eq5012}). 
With this, we find two equations
\begin{equation}
a(H\xi)^{\mbox{.}}=z^2 \tilde{c}^2\chi,
 \label{eq5014}
\end{equation}
\begin{equation}
a\dot{\chi}=z^{-2}H\nabla^2\xi,
 \label{eq5015}
\end{equation}
where
\begin{equation}
\xi=\frac{a\Phi}{4\pi G_{\rm N}H}, \quad \chi=\frac{\delta\theta}{\dot{\theta}} 
 \label{eq5016}
\end{equation}
and
\begin{equation}
z=\frac{a}{\tilde{c}}\sqrt{\frac{p+\rho}{\beta}} =
\frac{aH}{c_{\rm s}}\sqrt{\frac{(1-h^2/2)\varepsilon_1}{4\pi G_{\rm N}\gamma}} .
 \label{eq5017}
\end{equation}
In conformal time $\tau=\int dt/a$ equations (\ref{eq5014}) and (\ref{eq5015}) yield
a second order differential equation
\begin{equation}
v''-\tilde{c}^2 \nabla^2 v-\frac{z''}{z}v =0,
 \label{eq5018}
\end{equation}
where 
\begin{equation}
	v=z\chi .
\label{eq5500}
\end{equation}
The function $v$ is related to the gauge invariant  quantity  
\begin{equation}
 \zeta = \Phi+H\frac{\delta\theta}{\dot{\theta}},
\label{eq0021}
\end{equation}
introduced in Ref.\ \cite{garriga}. 
Indeed, using (\ref{eq5016}), (\ref{eq5500}), and 
(\ref{eq0021})
we have 
\begin{equation}
	v= z\left(\frac{\zeta}{H}-\frac{4\pi G_{\rm N}}{a}\xi  \right)\simeq \frac{z\zeta}{H}  ,
	\label{eq5019}
\end{equation}
where we have neglected  the second term in brackets 
being of higher order in $\varepsilon_i$ as shown in Appendix \ref{justify}.
 The quantity $\zeta$  measures the spatial curvature of comoving (or constant-$\theta$) hyper-surfaces. 

As usual, equation (\ref{eq5018}) is solved in momentum space where it reads
\begin{equation}
v_k''+\left(\tilde{c}^2k^2  -\frac{z''}{z}\right)v_k =0.
 \label{eq0032}
\end{equation}
The quantity $z''/z$ can be easily calculated up to the second order in $\varepsilon_i$.
However, as we have systematically  neglected the terms of order $\mathcal{O}(\varepsilon_i^2)$
it is consistent to keep only the dominant contribution to $z''/z$.
In the slow-roll regime one  can use  the relation
\cite{lidsey1}
\begin{equation}
\tau=- \frac{1+\varepsilon_1}{aH}+\mathcal{O}(\varepsilon_1^2),
\label{eq0047}
\end{equation}
which follows from the definition (\ref{eq3114}) expressed in terms of the conformal time.
Keeping the terms up to the first order one finds
\begin{equation}
\frac{z''}{z}=\frac{1}{\tau^2}(\nu^2-1/4) +\mathcal{O}(\varepsilon_i^2) ,
 \label{eq5020}
\end{equation}
where
\begin{equation}
\nu^2=\frac94+3\left(1+\frac{h^2}{2-h^2}\right)\varepsilon_1+3\varepsilon_2  .
 \label{eq5021}
\end{equation}

We look for
a solution to (\ref{eq0032}) which satisfies the positive frequency asymptotic limit
\begin{equation}
\lim_{\tau\rightarrow -\infty}v_k=\frac{e^{-i\tilde{c}k\tau}}{\sqrt{2\tilde{c}k}}.
 \label{eq0027}
\end{equation}
Then the  properly normalized solution to (\ref{eq5018}) which up to a phase agrees with (\ref{eq0027})
is
\begin{equation}
v_k=\frac{\sqrt{\pi}}{2}(-\tau)^{1/2} H_\nu^{(1)}(-\tilde{c}k\tau),
 \label{eq0049}
\end{equation}
where $H_\nu^{(1)}$ is the Hankel function of the first kind of rank $\nu$.
In the limit of the de~Sitter background all $\varepsilon_i$ vanish so $\nu=3/2$
in which case the solution to (\ref{eq0032}) is given by
\begin{equation}
v_k=\frac{e^{-i\tilde{c}k\tau}}{\sqrt{2\tilde{c}k}}\left(1-\frac{i}{\tilde{c}k\tau}\right) .
 \label{eq0051}
\end{equation}

Applying the standard canonical quantization \cite{mukhanov}  the field $v_k$ is promoted to an operator
 and the power spectrum of the field $\zeta_k=v_k/z$ is obtained  from the two-point correlation function
\begin{equation}
\langle\hat{\zeta}_k\hat{\zeta}_{k'}\rangle=\langle\hat{v}_k \hat{v}_{k'}\rangle/z^2=
(2\pi)^3 \delta(\mbox{\boldmath $k$}+\mbox{\boldmath $k$}')|\zeta_k|^2.
 \label{eq0063}
\end{equation}
The dimensionless spectral density
\begin{equation}
\mathcal{P}_{\rm S}(k)=\frac{k^3}{2\pi^2}|\zeta_k|^2=\frac{k^3H^2}{2\pi^2z^2}|v_k|^2 ,
 \label{eq5022}
\end{equation}
with $z$ given by (\ref{eq5017}),
characterizes the primordial scalar fluctuations.
Next,
  we evaluate the scalar spectral density at the horizon crossing, i.e., for  a wavenumber satisfying $k=aH$.
Following Refs.\ \cite{steer,hwang} we make use of the expansion of the Hankel function
in the limit $\tilde{c}k\tau\rightarrow 0$
\begin{equation}
 H_\nu^{(1)}(-\tilde{c}k\tau)\simeq -\frac{i}{\pi}\Gamma(\nu)\left(\frac{-\tilde{c}k\tau}{2}\right)^{-\nu},
 \label{eq0053}
\end{equation}
where the conformal time $\tau<0$ and $k$ is the comoving wavenumber.
Using this
we find  at the lowest order in $\varepsilon_1$ and $\varepsilon_2$
\begin{equation}
 \mathcal{P}_{{\rm S}} \simeq \frac{G_{\rm N} H^2 }{\pi c_{\rm s}\varepsilon_1  }
\frac{(\beta/\gamma)^{3/2}\gamma}{1-h^2/2}
\left[1-\left(2+2K-\ln \frac{\beta}{\gamma}\right)\varepsilon_1
-\left(2K-\ln \frac{\beta}{\gamma}\right)\left(\varepsilon_2+\frac{h^2}{2-h^2}\varepsilon_1\right)
\right]  ,
 \label{eq5023}
\end{equation}
 where $K=\gamma_{\rm E}-2+\ln 2 \simeq -0.730$ and 
 $\gamma_{\rm E}$ is the Euler constant.

It is worth comparing this expression with the standard $k$-inflation result \cite{garriga}
\begin{equation}
 \mathcal{P}_{{\rm S}} \simeq \frac{G_{\rm N} H^2 }{\pi c_{\rm s}\varepsilon_1 }
 \left[1-2\left(1+K\right)\varepsilon_1-K \varepsilon_2\right] .
\label{eq3007}
\end{equation}
In the regime where $h^2\ll \varepsilon_i$, $\beta \rightarrow 1$, and $\gamma \rightarrow 1$, we recover
the standard result apart from a difference by  a factor of 2 in the $\varepsilon_2$ correction
in square brackets.
 The reason for this discrepancy is due to the linear dependence of $z$ on $\varepsilon_1$
as opposed to $\sqrt{\varepsilon_1}$ dependence in the standard case.
Although the field equations of MGB gravity become identical to the field equations of general 
relativity (GR) in the limit $h\to  0$,  MGB gravity is appreciably  different from GR for small $\varepsilon_1$ 
(when $\rho$ is expected to be large).
 Hence, GR need not have been recovered if one first expands in $\epsilon_1$ and then takes the limit $h \to 0$, as in the mentioned regime $h^2\ll \varepsilon_i$. 
 
In the ultra-slow-roll regime where $\varepsilon_i\ll h^2$
we find a substantial  enhancement with respect to the standard result
by the factor $\gamma\simeq (4\eta+3) h^2/(18\eta\varepsilon_1)$.

\subsection{Tensor perturbations}
\label{tensor2}

 The tensor perturbations are related to the production of gravitational waves
 during inflation.
 The metric perturbations
 are defined as
 \begin{equation}
 ds^2= dt^2-a^2(t)\left(\delta_{ij}+h_{ij}\right)dx^idx^j,
 \label{eq0054}
\end{equation}
where $h_{ij}$ is traceless and transverse.
Inserting the metric components in the field equations (\ref{eq:mod}) yields an equation for
the perturbation $h_{ij}$ which we derive in
appendix \ref{app-tensor}.
Assuming as usual no contribution from matter we write  the equation  for $h_{ij}$, Eq.\ (\ref{eq5200}),
in the form
\begin{equation}
A \ddot{h}_{ij}+ B 3H\dot{h}_{ij} + DH^{2}h_{ij}
-C\frac{\nabla^{2}h_{ij}}{a^{2}}=0 ,
 \label{eq3211}
\end{equation}
where $A$, $B$, $C$, and $D$ are functions of $H$ and its derivatives which
we can be expressed in terms of $\varepsilon_i$.
Using (\ref{eq3214}) and (\ref{eq3114}) we find
\begin{equation}
A = 1+ \frac{h^{2}}{6\varepsilon_{1}}(1+\varepsilon_{2}),
 \label{eq3219}
\end{equation}
\begin{equation}
B = 1+ \frac{h^{2}}{6\varepsilon_{1}}\left( 1-\frac{2}{3}\varepsilon_{1}
+\frac{2}{3}\varepsilon_{2} +\frac{8}{3}\varepsilon_{1}^{2} - \varepsilon_{2}^{2}
- \frac{2}{3}\varepsilon_{1}\varepsilon_{2} + \frac{1}{3}\varepsilon_{2}\varepsilon_{3} \right),
 \label{eq3220}
\end{equation}
\begin{equation}
C = 1+ \frac{h^{2}}{6\varepsilon_{1}}\left( 1 - \varepsilon_{2}^{2} - \varepsilon_{1}\varepsilon_{2}
+ \varepsilon_{2}\varepsilon_{3} \right),
 \label{eq3221}
\end{equation}
\begin{equation}
D = \frac{h^{2}}{3 \varepsilon_{1}}\varepsilon_{2} .
\end{equation}
We now proceed as in section \ref{scalar} and divide (\ref{eq3211}) by $A$
\begin{equation}
\ddot{h}_{ij} +\frac{B}{A}3H \dot{h}_{ij}+\frac{D}{A}H^2h_{ij}- \frac{C}{A}\frac{\nabla^2 h_{ij}}{a^2}=0 .
 \label{eq3222}
\end{equation}
At the beginning and at the end of inflation the coefficient $D/A$ tends to zero
and $B/A$ and $C/A$ both tend to unity.
At quadratic order in $\varepsilon_i$ we find
\begin{equation}
\frac{B}{A} = 1-\frac{2}{3}\varepsilon_{1} - \frac{1}{3}\varepsilon_{2}
+ \left(\frac{8}{3} +\frac{4(h^{2}-9)}{h^{4}}  \right)\varepsilon_{1}^{2}
- \frac{2}{3}\varepsilon_{2}^{2} +\frac{2}{h^{2}}\varepsilon_{1}\varepsilon_{2}
+\frac{1}{3}\varepsilon_{2}\varepsilon_{3},
 \label{eq3223}
\end{equation}
\begin{equation}
\frac{C}{A} = 1-\varepsilon_{2} - \frac{36}{h^{4}}\varepsilon_{1}^{2}
+ \left( -1+\frac{6}{h^{2}} \right)\varepsilon_{1}\varepsilon_{2}+\varepsilon_{2}\varepsilon_{3},
 \label{eq3224}
\end{equation}
\begin{eqnarray}
\frac{D}{A} = 2\varepsilon_{2} - 2\varepsilon_{2}^{2} - \frac{12}{h^{2}}\varepsilon_{1}\varepsilon_{2} .
\end{eqnarray}
Now we proceed by solving equation (\ref{eq3222})
in the usual way.
Keeping the linear order in $\varepsilon_i$, Eq.\ (\ref{eq3222}) becomes
\begin{eqnarray}
\ddot{h}_{ij}+  H(3 -2\varepsilon_{1}-\varepsilon_{2})\dot{h}_{ij} + 2H^{2}\varepsilon_{2}h_{ij} -
 (1-\varepsilon_{2})\frac{\nabla^{2}h_{ij}}{a^{2}} = 0.
\end{eqnarray}
To solve this one uses
the standard Fourier decomposition in conformal time $\tau$
\begin{equation}
h_{ij}(\tau,\mbox{\boldmath $x$})= \frac{1}{(2\pi)^3}\int d^3k  e^{i\mbox{\scriptsize\boldmath $kx$}}
\sum_s h_k^s(\tau)e^s_{ij}(k),
 \label{eq0056}
\end{equation}
where the polarization tensor
$e^s_{ij}$  satisfies
$k^i e^s_{ij}=0$, and $e^s_{ij} e^{s'}_{ij}=2\delta_{ss'}$
with comoving wavenumber $k$ and two polarizations $s=+,\times$.
The amplitude $h_k^s(\tau)$ then satisfies
\begin{equation}
  h_{k}'' + 2 aH h_{k}' - (2\varepsilon_{1}+\varepsilon_{2})aH h_{k}' +2 a^{2}H^{2}\varepsilon_{2}h_{k} +
(1-\varepsilon_{2})k^{2}h_{k} = 0,
 \label{eq3226}
\end{equation}
where we have suppressed the dependence on $s$ for simplicity  bearing in mind
that we have to sum over two polarizations in the final expression.
Note that the third term may be neglected as it is suppressed by a factor $\mathcal{O}(\varepsilon_i^2)$
with respect to the last term.
This may be seen by estimating the ratio $aH h_k'/(k^2 h_k)$. Employing the trick
(\ref{eq5300}) we estimate $\dot{h}_k\simeq  \mathcal{O}(\varepsilon_1)H h_k$
and find
\begin{equation}
\frac{a H h_k'}{k^2 h_k}\simeq \frac{a^2 H^2 }{k^2}\mathcal{O}(\varepsilon_1)
 \simeq \mathcal{O}(\varepsilon_1).
 \label{eq3227}
\end{equation}
In the second equality we have used the value
\begin{equation}
k\simeq aH
 \label{eq3228}
\end{equation}
near the horizon crossing.
Thus, neglecting the suppressed term and introducing a canonically normalized amplitude
\begin{equation}
v_k=\frac{a}{16\pi G_{\rm N}}h_k
 \label{eq0058}
\end{equation}
we obtain the equation
 \begin{equation}
{v_k}''+\left(k^2  -\frac{a''}{a}+a^2H^2 \varepsilon_2\right)v_k =0.
 \label{eq3229}
\end{equation}
This equation is of the same form as (\ref{eq0032}) with $c_{\rm s}=1$ and $z''/z$ replaced by $a''/a-a^2H^2 \varepsilon_2$.
As before, using the relations (\ref{eq0047}) and (\ref{eq3228}) we find a
properly normalized solution
\begin{equation}
v_k=\frac{\sqrt{\pi}}{2}(-\tau)^{1/2} H_\nu^{(1)}(-k\tau),
 \label{eq3230}
\end{equation}
with
\begin{equation}
\nu^2=9/4+3 \varepsilon_1-\varepsilon_2.
 \label{eq3231}
\end{equation}
The spectral density of the primordial tensor fluctuations
is then given by
\begin{equation}
\mathcal{P}_{\rm T}(k)=\frac{k^3}{\pi^2}|h_k|^2=
\frac{k^3}{\pi^2}\left|\frac{16\pi G_{\rm N}}{a}v_k\right|^2 ,
 \label{eq0065}
\end{equation}
with $v_k$ given by (\ref{eq3230}).
Then, at the horizon crossing, using the approximation (\ref{eq0053}) we find
\begin{equation}
 \mathcal{P}_{{\rm T}} \simeq \frac{16 G H^2 }{\pi}
 \left[1-2\left(1+K\right)\varepsilon_1+\frac23 K\varepsilon_2\right].
\label{eq3232}
\end{equation}

\section{Scalar spectral index and tensor to scalar ratio}
\label{index}

The scalar spectral index $n_{\rm S}$ and tensor to scalar ratio $r$ are given by
\begin{equation}
 n_{\rm S}-1= \frac{d\ln \mathcal{P}_{{\rm S}}}{d\ln k}\simeq\frac{1}{H(1-\varepsilon_1)}\frac{d\ln \mathcal{P}_{{\rm S}}}{dt},
 \label{eq3006}
 \end{equation}
 \begin{equation}
 r=\frac{\mathcal{P}_{\rm T}}{\mathcal{P}_{{\rm S}}},
 \label{eq3005}
 \end{equation}
 where $\mathcal{P}_{{\rm S}}$ and $\mathcal{P}_{\rm T}$ are
 evaluated at the horizon crossing.
 The second equality in  (\ref{eq3006}) is obtained with the help of (\ref{eq0047})
and  the horizon crossing relation $k=a H$.

To be consistent with our approximation, in the calculation of $n_{\rm S}$ and $r$
 we keep only the lowest order corrections to the leading term.
From (\ref{eq5023}) and (\ref{eq3232})
we find at linear order
\begin{equation}
	n_{\rm S}=1- 2\left(1+\frac{2}{2-h^2}  \right)\varepsilon_1-2\varepsilon_2 +\mathcal{O} (\varepsilon_i^2)
\label{eq131}
\end{equation}
and
\begin{equation}
r=\frac{144(2-h^2)}{h^2}\frac{\eta(4\eta+3)^{1/2}}{(1-\eta)^{3/2}} \varepsilon_1^2 
	\left[1+ a\varepsilon_1+b\varepsilon_2+\mathcal{O} (\varepsilon_i^2)
\right],
 \label{eq130}
\end{equation}
where 
\begin{equation}
a=\frac{2h^2}{2-h^2}K+\frac{2}{2-h^2}\ln \frac{4\eta+3}{1-\eta}
-\frac{9(\eta^2+11\eta+9)}{h^2(4\eta+3)(1-\eta)}+
\frac{3(7\eta^2+43\eta+27)}{2(4\eta+3)(1-\eta)}
+\frac{2(2-h^2)}{3(4-h^2)}\frac{p p_{,XX}}{p_{,X}^2} ,
\label{eq0130}
\end{equation}
\begin{equation}
b=\frac83 K + 
\ln \frac{4\eta+3}{1-\eta}+\frac{3\eta(3\eta+4)}{2(4\eta+3)(1-\eta)}.
\label{eq1=0131}
\end{equation}

For comparison, it is worth quoting the  results
we have obtained
in holographic cosmology (HC) \cite{bertini}
\begin{equation}
\left. r\right|_{\rm HC}=16 \varepsilon_1 \left[1+K\varepsilon_2
+\frac{2(2-h^2)}{3(4-h^2)}\frac{p p_{,XX}}{p_{,X}^2}\varepsilon_1\right]
 \label{eq128}
\end{equation}
and
\begin{eqnarray}
&& \left. n_{\rm S}\right|_{\rm HC}=1-2\varepsilon_1-\varepsilon_2
-\left(2+\frac{8h^2}{3(4-h^2)^2}\frac{p p_{,XX}}{p_{,X}^2}\right)\varepsilon_1^2
\nonumber\\
&& -\left(3+2K +\frac{2(2-h^2)}{3(4-h^2)}\frac{p p_{,XX}}{p_{,X}^2}\right)\varepsilon_1\varepsilon_2
- K \varepsilon_2\varepsilon_3 ,
 \label{eq129}
\end{eqnarray}
which, in the limit $h^2\rightarrow 0$, coincide with the results obtained in the standard $k$-inflation \cite{hwang}
in general relativistic cosmology.
Clearly, there is a significant deviation from the standard cosmology:
the leading term in (\ref{eq130}) is suppressed by a factor 
$$
\frac{9(2-h^2)}{h^2}\frac{\eta(4\eta+3)^{1/2}}{(1-\eta)^{3/2}} \varepsilon_1
$$
compared with  (\ref{eq128}) and the first order corrections
to $n_{\rm S}$ in (\ref{eq131}) are enhanced roughly by a factor of 2 compared with
(\ref{eq129}).
It is interesting to note that a similar suppression of the leading term in $r$ is obtained in
a recent model \cite{odintsov3} based on $f(R)$ gravity.

Our analysis and the results obtained so far have been basically model independent.
To make a comparison with observation,  e.g., to plot $r$ versus $n_{\rm S}$,
we need to specify a model.
 A  model which can be easily treated
is the  tachyon condensate \cite{sen} which has been extensively studied 
 in the context of inflation
 \cite{steer,fairbairn,frolov,shiu1,sami,shiu2,kofman,cline,salamate2018,barbosa2018,dantas2018}.
Tachyon models are of particular interest as in these models inflation 
is driven by the tachyon field originating in M or string theory.
The basics  of the tachyon condensation
 are contained in an effective field theory  \cite{sen}
with Lagrangian of the Dirac-Born-Infeld (DBI) form 
\begin{equation}
\mathcal{L}_{\rm DBI} = -\ell^{-4} V(\theta/\ell)
\sqrt{1-g^{\mu\nu}\theta_{,\mu}\theta_{,\nu}}  .
 \label{eq000}
\end{equation}
The dimensionless potential $V$ is a positive function of $\theta$ with a unique local maximum 
at $\theta=0$ and a global minimum at $\theta=\infty$ at which $V$ vanishes. 
 A simple potential which  satisfies the above requirements 
	is the exponential potential
\begin{equation}
V=e^{-\omega\theta} ,
\end{equation}
where $\omega$ is a parameter of dimension of mass. 
This potential has been studied in Ref.\ \cite{bilic3} in the context of holographic braneworld inflation.

For the tachyon model with exponential potential we have
\begin{equation}
\frac{p p_{,XX}}{p_{,X}^2}= -1
\end{equation}
and
\begin{equation}
\varepsilon_2
\simeq 2\varepsilon_1\left(1-\frac{2h^2}{(2-h^2)(4-h^2)}\right) ,
 \label{eq127}
\end{equation}
where $h\equiv \ell H$ is a monotonously decreasing function of time according to the second Friedmann equation (\ref{eq0102}).
Owing to (\ref{eq125}), the initial value of $h$  
at $t=0$
must satisfy the restriction
\begin{equation}
h_{\rm i}\leq \sqrt2  .
\label{eq5125}
\end{equation}
 The choice of $\omega$ and $h_{\rm i}$ affects the e-fold number defined as
\begin{equation}
N=\int_0^{t_0} H dt ,
\label{eq6025}
\end{equation}
where $t_0$ is the duration of the slow-roll regime
fixed by the requirement $\varepsilon_1(t_0)=1$.
Hence, $N$ is an implicit function of $\omega$ and 
  $h_{\rm i}$ only.
\begin{figure}[ht]
\begin{center}
\includegraphics[width=\textwidth ]{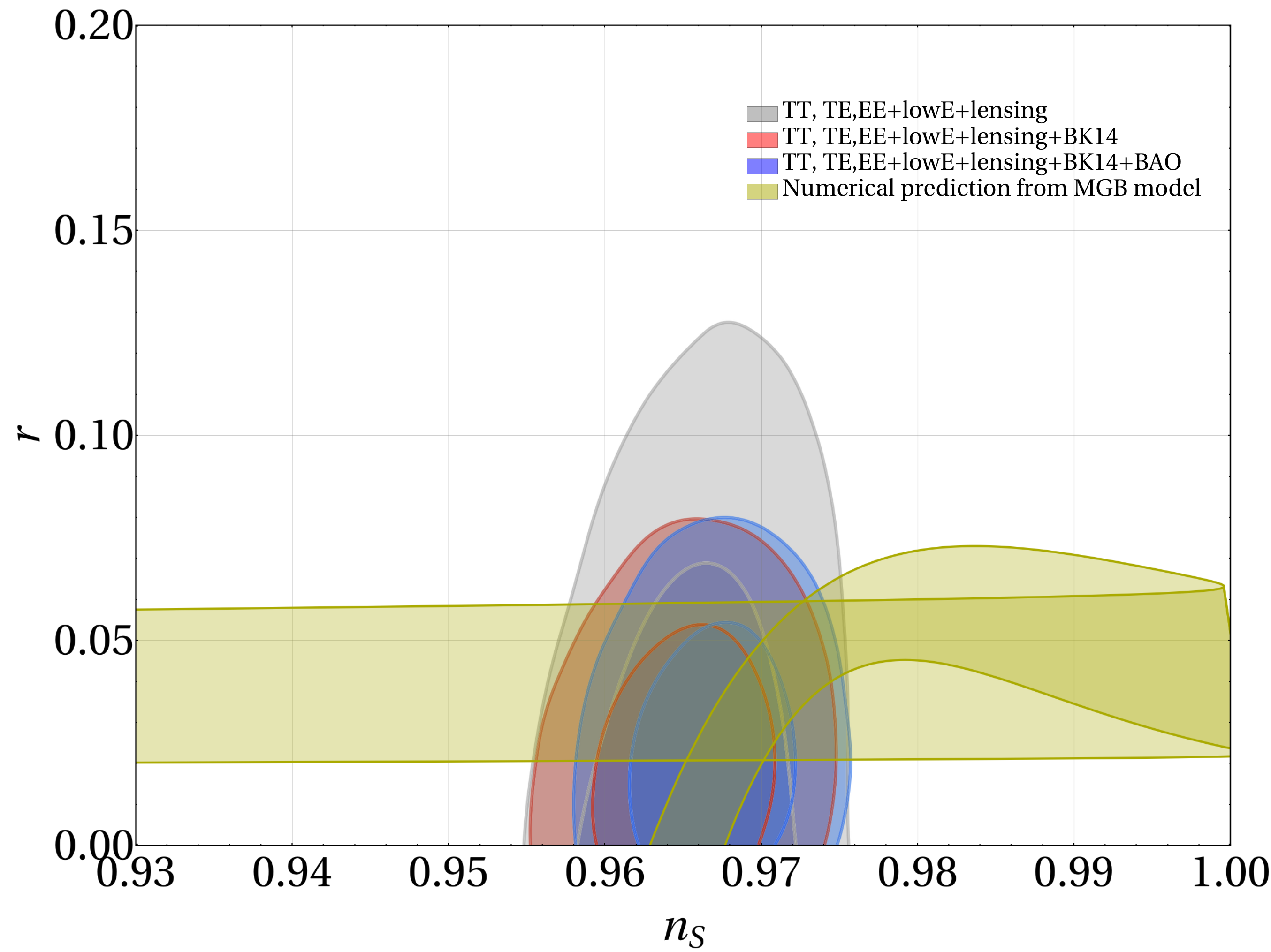}
\caption{$r$ versus $n_{\rm s}$ diagram with observational constraints
from Ref.\  \cite{planck2018b}. 
	The yellow shaded strips  represents theoretical predictions
for fixed  $h_{\rm i}=1$ and $\eta=0.5$ and $N$ ranging from 50 (top boundary of the strips) to 90
(bottom boundary of the  strips).}
\label{fig1}
\end{center}
\end{figure}
\begin{figure}[ht]
	\begin{center}
\includegraphics[width=\textwidth ]{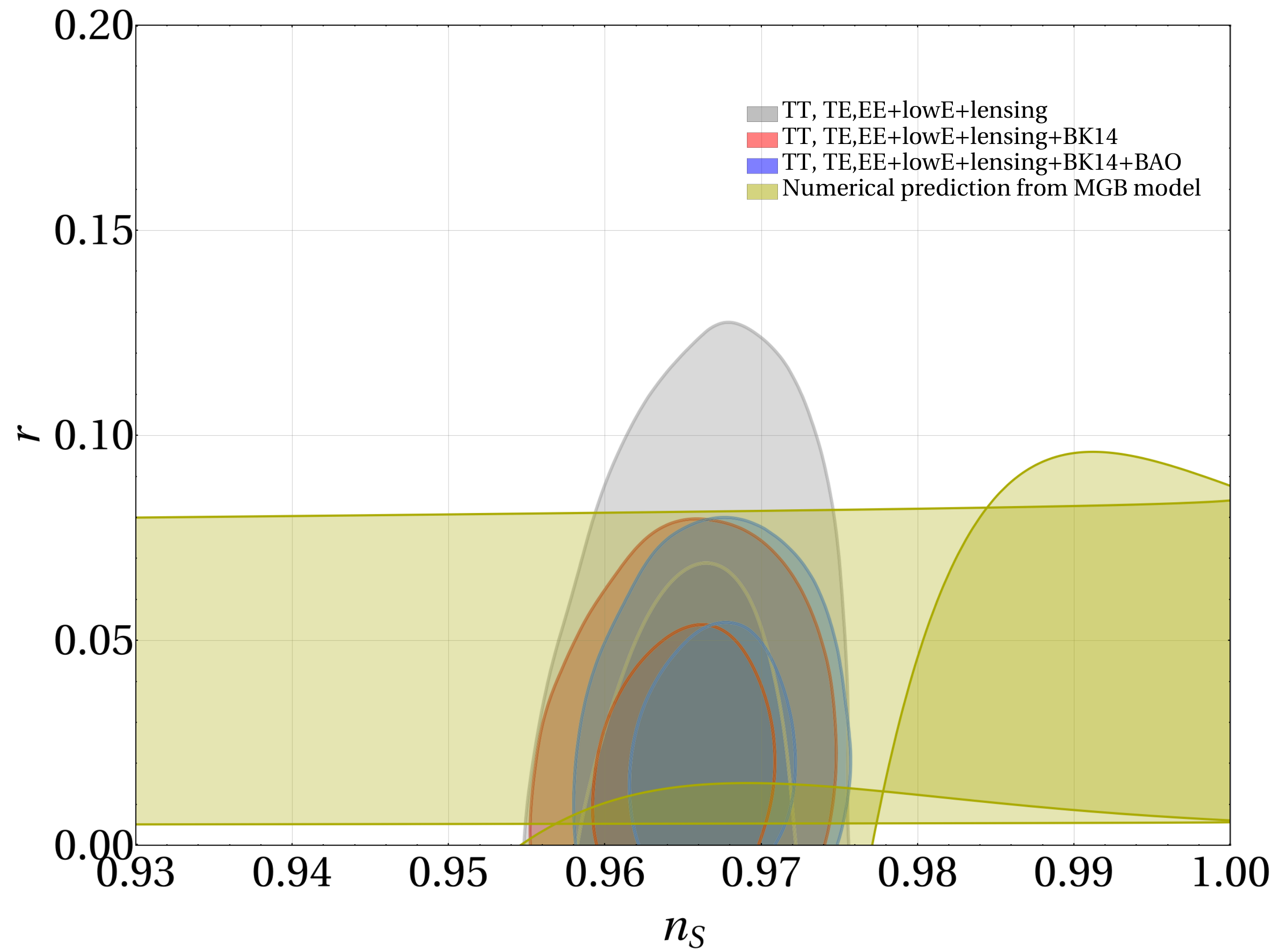}
\caption{$r$ versus $n_{\rm s}$ diagram with observational constraints
from Ref.\  \cite{planck2018b}.
 	The yellow shaded strips  represents theoretical predictions
for fixed  $h_{\rm i}=1$ and $N=75$ and $\eta$ ranging from 0.2 (bottom boundary of the strips) to 0.7
(top boundary of the  strips).}
\label{fig2}
\end{center}
\end{figure}
Our basic results (\ref{eq131}) and (\ref{eq130}) also depend on the
slip parameter $\eta$ which, according to (\ref{eq:off-diag}), is generally a function of $k$ and $t$. Our estimate at the horizon crossing allows us to assume that
$\eta$, being a smooth function of $t$, is roughly constant
during the slow-roll regime.
For simplicity, in the following, we will treat $\eta$ as a free parameter with values in the interval $0 < \eta < 1$.

 The numerical calculations proceed  as follows.  For chosen $\omega$, $h_i$, and $\eta$ we evolve our background with $t$ ( $0<t<t_0$) to get $h(t)$
and $\varepsilon_1(t)$ and produce $h=h(\varepsilon_1)$ as a parametric function. 
For each $\eta$ and initial $h_{\rm i}$ the value of $t_0$ is fixed by $\varepsilon_1(t_0)=1$. 
This gives $N$ as a function of $\omega$ for each fixed  $\eta$ and $h_{\rm i}$. This function can be numerically inverted to obtain $\omega=\omega(N)$.
In this way, for fixed $h_i$ and $\eta$ we can produce a set of curves $r=r(n_{\rm S})$ each labeled by a value of $N$. Similarly, for fixed $N$ and $h_i$ we can produce another set of curves $r=r(n_{\rm S})$ each labeled by a value of $\eta$.

In Fig.\ \ref{fig1} 
the theoretical plot for fixed initial value  $h_{\rm i}= 1$ and $\eta=0.5$
 is superimposed on the observational constraints taken from the Planck Collaboration 2018  \cite{planck2018b}. 
The parameter $\omega$ is allowed to vary so that the e-fold number $N$ varies in the range $50\leq N \leq 90$.
The central  point where the lines $N=50$ and $N=90$
cross corresponds to
 $n_{\rm S}=0.965$ and $r= 0.0205$, in
excellent agreement with observations.
Similarly,   the theoretical plot for varying slip parameter $\eta$ in the range $0.2\leq \eta \leq 0.7$ is presented in Fig.\ \ref{fig2}. In this plot, the e-fold number $N=75$ and initial  $h_{\rm i}= 1$ are kept fixed. 
Both figures demonstrate that there exist a reasonable set of parameters for which the theoretical prediction is in good agreement with observations.  
\section{Summary and  conclusions}
\label{conclude}

We have studied the early universe  cosmology by means of a modified gravity model in which  the gravity action consists of  a function of  the Ricci scalar and the Gauss-Bonnet invariant in addition to the Einstein-Hilbert term. The field equations are obtained by making use of the  scalar-tensor representation of the action. We have specified the functional form of the action so that the modified Friedmann equations have the same form as those obtained in the holographic cosmology scenario.
Furthermore, we have 
 developed the formalism for calculating  cosmological perturbations for scalar and tensor modes with matter represented by a general $k$-essence field theory. Using  this, we have derived the scalar and tensor power spectra and calculated  the scalar-to-tensor ratio $r$ and spectral index $n_{\rm S}$.

To confront our model with observations we have calculated  $n_{\rm S}$ and $r$ for a particular tachyon type $k$-essence  with exponential potential.
Our numerical results (see fig.~\ref{fig1}) show that the predictions of the MGB model are consistent with the Planck observational constraints.
	In this comparison we have fixed the initial expansion rate to $h_{\rm i}=1$ and the slip parameter to $\eta=0.5$, and the only remaining free parameter, 
the parameter $\omega$ in the potential, has been allowed to vary in such a way  that the e-fold number $N$ varies in the physically acceptable range $50\leq N \leq 90$.

One of our aims has been to compare the MGB model with the holographic cosmology. We have recently studied inflation within the latter model \cite{bertini}, where a modest departures from standard GR with $k$-essence has been found: for an inflationary scenario with $k$-essence, the differences in the power spectrum are only found at the second order in the slow-roll parameters (due to a change in the speed of sound).  The expression   (\ref{eq130}) shows that the tensor-to-scalar ratio $r$ departs from zero only at second order in the slow parameters in contrast to the holographic cosmology or the standard GR cosmology  with $k$-essence where the departure from zero is at first order.

\section*{Acknowledgments}
N.R.~Bertini thanks CAPES (Brazil) for support. The work of N.~Bili\'c  has been partially supported by
the European Union through the European Regional Development Fund - the Competitiveness and
Cohesion Operational Programme (KK.01.1.1.06) and by the ICTP - SEENET-MTP project NT-03 Cosmology - Classical and Quantum Challenges. 
D.C.~Rodrigues thanks  CNPq (Brazil) and FAPES (Brazil) for partial support.
This study was financed in part by the
{\it Coordena\c{c}\~ao de Aperfei\c{c}oamento de Pessoal de N\'ivel Superior - Brasil} (CAPES) - Finance Code 001.

\appendix

	\section{Justification for the approximations made}
	\label{justify}

	Here we justify the approximations of neglecting
	the term $\Phi/\eta$ in Eq.\ (\ref{eq5012})
	and the term $4\pi G_N \xi/a$ in Eq.\ (\ref{eq5019}).
	First, we estimate the magnitude of the second term on the right-hand side of (\ref{eq5012}) in momentum
	space.
	Using the second Friedmann equation,  the definition of $\varepsilon_1$, and the approximate value in the ultra slow-roll regime
	$\gamma\simeq (4\eta+3) h^2/(18\eta\varepsilon_1)$ we  find
	\begin{equation}
		\frac{c_{\rm s}^2\gamma k^2 \Phi}{4\pi G_{\rm N}a^2(p+\rho)} =
		\frac{(4\eta+3)h^2c_{\rm s}^2 k^2 \Phi}{18\eta a^2H^2\varepsilon_1^2(1-h^2/2)} .
		\label{eq5024}
	\end{equation}
	To make an  order of magnitude estimate we can use the value $k\simeq aH/c_{\rm s}$ near the acoustic horizon crossing.
	With this we find
	\begin{equation}
		\frac{c_{\rm s}^2\gamma k^2 \Phi}{4\pi G_{\rm N}a^2(p+\rho)} \simeq
		\frac{ (4\eta+3)h^2 \Phi}{9\eta(2-h^2)\varepsilon_1^2}\gg \Phi,
	\end{equation}
	which justifies the approximation of neglecting the term $\Phi/\eta$ in Eq.\ (\ref{eq5012}).
	
	Next, we estimate the order of magnitude of $4\pi G_N \xi/a$ in comparison
	with $\chi\simeq\zeta/H$ in Eq.\ (\ref{eq5019}).
	Applying (\ref{eq5300}) to Eq.\ (\ref{eq5014}) we find
	\begin{equation}
		\mathcal{O}(\varepsilon_1)H^2\xi\simeq \frac{1}{a}z^2 \tilde{c}^2\chi=
		\frac{aH^2 9\eta(2-h^2)\varepsilon_1^2}{4\pi G_{\rm N}(1-\eta)h^2}\chi
	\end{equation}
	and hence
	\begin{equation}
		\frac{4\pi G_{\rm N}}{a}\xi\simeq \frac{9\eta(2-h^2)}{(1-\eta)h^2}\frac{\varepsilon_1^2}{\mathcal{O}(\varepsilon_1)}\chi \ll \chi .
	\end{equation}
	This justifies the approximation made to obtain the second equality in Eq.\ (\ref{eq5019}).

\section{Scalar perturbations}
\label{app-scalar}
In this appendix we explicitly derive the equations  for linear order scalar mode perturbations in the Newtonian gauge
for the MGB model \eqref{eq:2}, \eqref{eq:3}.
We apply the usual metric formalism and assume that Christoffel symbols
are related to the metric through the Levi-Civita connection.
Using the perturbed components of Christoffel  symbols in the expressions for Ricci and Riemann tensor
one can obtain all the perturbed quantities of the equation \eqref{eq:mod}.
For the sake of completeness, in Sec.\ \ref{sec:pgq}  we
provide the expressions for all these geometric quantities
together with perturbed Ricci scalar and Gauss-Bonnet invariant.
In Sec.\ \ref{sec:pfe} we derive the expressions for the auxiliary fields $\psi_{1}$ and $\psi_{2}$,
and the final equations for the scalar perturbations.

\subsection{Perturbed geometric quantities}\label{sec:pgq}
For the line element  \eqref{eq0013}, the metric components are
\begin{equation}
g_{00} = 1+ 2\Psi,
\label{eq:pertg00}
\end{equation}
\begin{equation}
g_{ij} = - a^{2}(t)(1-2\Phi)\delta_{ij}.
\label{eq:pertgij}
\end{equation}
Plugging the above metric components into the Cristoffel symbols we find
\begin{eqnarray}
\Gamma^{0}_{00} = \dot{\Psi}, \qquad \Gamma^{i}_{00} = \frac{\delta^{ij}}{a^{2}}\partial_{j}\Psi, \qquad
\Gamma^{0}_{ij} = a^{2}[H- 2 H(\Phi + \Psi) - \dot{\Phi}]\delta_{ij},\nonumber
\\
\Gamma^{i}_{0j} = (H - \dot{\Phi})\delta^{i}_{j}, \qquad \Gamma^{i}_{jk} =
\delta^{i}_{k}\partial_{k}\Psi + \delta^{i}_{j}\partial_{k}\Psi - \delta^{il}\delta_{jk}\partial_{l}\Psi. \qquad
\end{eqnarray}
The components of the Riemann and Ricci tensor are
\begin{align}
R_{0i0j}  & = a^{2}(H^{2}+\dot{H})\delta_{ij} - \partial_{i}\partial_{j}\Psi
- a^{2}\delta_{ij}\big[2(H^{2}+\dot{H})\Phi + H (2\dot{\Phi}+\dot{\Psi})+ \ddot{\Phi}\big],
\\
\nonumber
\\
R_{ijkl}  &=  - a^{4}H^{2}(\delta_{ik}\delta_{lj}- \delta_{il}\delta_{kj})
+ a^{2}(\delta_{li}\delta^{n}_{k}\delta^{m}_{j} - \delta_{jl}\delta^{n}_{k}
\delta^{m}_{i} - \delta_{ki}\delta^{m}_{l}\delta^{n}_{j}
+\delta_{kj}\delta^{m}_{l}\delta^{i}_{n} )\partial_{m}\partial_{n}\Phi
\nonumber\\
	 &+ (\delta_{ik}\delta_{jl} - \delta_{il}\delta_{jk})\big[2a^{4}H^{2}(\Psi+ 2\Phi) +2 a^{4}H \dot{\Psi}\big],
\\
\nonumber
\\
R_{00}  &=   -3 (H^{2}+\dot{H})+ \frac{1}{a^{2}}\nabla^{2}\Psi + 3H (2\dot{\Phi}+\dot{\Psi}) + 3\ddot{\Phi},
\\
\nonumber
\\
 R_{0i} &=   2\partial_{i}(\dot{\Phi}+H\Psi),
\\
\nonumber
\\
R_{ij} &=   a^{2}(3H^{2}+\dot{H})\delta_{ij} + \partial_{i}\partial_{j}(\Phi - \Psi)
+ \delta_{ij} \nabla^{2}\Phi\nonumber
\\
  &- a^{2}\delta_{ij}\big[ 2(3H^{2}+\dot{H})(\Phi+\Psi) + H(6\dot{\Phi}+\dot{\Psi}) + H\ddot{\Phi}\big].
\end{align}
With the help of the above quantities the perturbed Ricci scalar and Gauss-Bonnet invariant can be calculated yielding
\begin{equation}
R = -6 (2 H^{2}+ \dot{H}) + \frac{2}{a^{2}}\nabla^{2}\Psi - \frac{4}{a^{2}}\nabla^{2}\Phi + 6
\big[ 2(2H^{2}+\dot{H})\Psi + H(4\dot{\Phi}+\dot{\Psi}) + \ddot{\Phi} \big]
\end{equation}
and
\begin{eqnarray}
&& {\cal G} = 24 H^{2}(H^{2}+ \dot{H}) + \frac{16}{a^{2}}(H^{2}+\dot{H})\nabla^{2}\Phi
 - \frac{8}{a^{2}}H^{2}\nabla^{2}\Psi - 96H^{2}(H^{2}+\dot{H})\Psi
\nonumber\\
&& -48 H(2H^{2}+\dot{H})\dot{\Phi} - 24 H^{3}\dot{\Psi} - 24H^{2}\ddot{\Phi}.
\end{eqnarray}

\subsection{Perturbed field equations}\label{sec:pfe}
 Scalar modes induce fluctuations in all quantities in \eqref{eq:mod}. 
Inserting the scalar perturbations of the metric in the Newtonian gauge
and the field perturbations $\psi_{a} \rightarrow \psi_{a}(t)+ \delta\psi_{a}$,   
the components of the linear part of Eq. \eqref{eq:mod} become
\begin{align}
	& \frac{1}{a^{2}}\big[  \nabla^{2}\delta\psi_{1} - 4H^{2}\nabla^{2}\delta\psi_{2} - 2(\psi_{1}
- 4 H\dot{\psi}_{2})\nabla^{2}\Phi  \big] + 3(H^{2}+\dot{H})(\delta\psi_{1} - 4H^{2}\delta\psi_{2})
\nonumber\\
& 
+ 6H(\dot{\psi}_{1}+ H(\psi_{1}-8H\dot{\psi}_{2}))\Psi - 3H\dot{\delta\psi}_{1}  + 12 H^{3}\dot{\delta\psi}_{2}
\nonumber\\
&  + 3(\dot{\psi}_{1}+2H(\psi_{1}-6H\dot{\psi}_{2}))\dot{\Phi}
= 8\pi G \delta T^{0}_{0},
	\label{eq:130}
\end{align}
\begin{align}
	&\partial_{i}\big[  -2(\psi_{1}-4H\dot{\psi}_{2})\dot{\Phi} + (12 H^{2}\dot{\psi}_{2} - 2H\psi_{1} - \dot{\psi}_{1})\Psi
- H\delta\psi_{1}   + 4 H^{3}\delta\psi_{2}
+\dot{\delta\psi}_{1}
\nonumber\\
&	 -4H^{2}\dot{\delta\psi}_{2} \big] = 8\pi G\delta T^{0}_{i},
\label{eq:131}
\end{align}
\begin{align}
	&\frac{1}{a^{2}}\big[ \nabla^{2}\delta\psi_{1} - 4(H^{2}+\dot{H})\nabla^{2}\delta\psi_{2} - (\psi_{1}-4\ddot{\psi}_{2})\nabla^{2}\Phi + (\psi_{1}-4H\dot{\psi}_{2})\nabla^{2}\Psi \big]\delta^{i}_{j}
\nonumber\\
&+ \Bigg[(3H^{2}+\dot{H})\delta\psi_{1} - 12H^{2}(H^{2}+\dot{H})\delta\psi_{2}
\nonumber\\
&+2\big(  2\psi_{1}\dot{H} - 16H^{2}\dot{\psi}_{2} + 2 H (\dot{\psi}_{1} - \dot{H}\dot{\psi}_{2}) + \ddot{\psi}_{2} + H^{2}(3\psi_{1} - 8\ddot{\psi}_{2})   \big)\Psi
\nonumber\\
&- 2H \dot{\delta\psi}_{1}  + 8H(H^{2}+\dot{H})\dot{\delta\psi}_{2} +  2\big(  \dot{\psi}_{1} - 12H^{2}\dot{\psi}_{2} - 4 \dot{H}\dot{\psi}_{2} + H(3\psi_{1} - 4\ddot{\psi}_{2})  \big)\dot{\Phi}
\nonumber\\
&+ \big(  \dot{\psi}_{1}+ 2 H ( \psi_{1} - 6H\dot{\psi}_{2}) \big)\dot{\Psi} + 2 (\psi_{1} - 4H\dot{\psi}_{2})\ddot{\phi} - \ddot{\delta\psi}_{1} - 4H^{2}\ddot{\delta\psi}_{2}
\Bigg]\delta^{i}_{j}
\nonumber\\
&- \frac{1}{a^{2}}\partial^{i}\partial_{j}\delta\psi_{1} + \frac{4}{a^{2}}(H^{2}+\dot{H})\partial^{i}\partial_{j}\delta\psi_{2} + \frac{1}{a^{2}}(\psi_{1}-4\ddot{\psi}_{2})\partial^{i}\partial_{j}\Phi
\nonumber\\
&- \frac{1}{a^{2}}(\psi_{1} - 4H\dot{\psi}_{2})\partial^{i}\partial_{j}\Psi = 8\pi G \delta T^{i}_{j}.
 \label{eq:132}
\end{align}

The perturbed auxiliary fields are functions of the invariants $X^a$ and hence
\begin{equation}
\psi_{b}(X^{a}+ \delta X^{a}) \approx \psi_{b}(X^{a})
	+ \frac{\partial\psi_{b}}{\partial X^{c}}\delta X^{c}.
\end{equation}
Then, from \eqref{eq5400} it follows
\begin{equation}
	\delta\psi_{b} = \frac{\partial^{2}F}{\partial X^{c}\partial X^{b}}\delta X^{c}.
\end{equation}
Using \eqref{eq:3} together with the spatially flat background metric
the linear parts of the auxiliary fields become
\begin{equation}
	\delta\psi_{1} = \frac{\ell^{2} H^{4}(2H^{2}+3\dot{H})}{36\dot{H}^{2}}\delta R
	+ \frac{\ell^{2} H^{2}(H^{2}+\dot{H})}{72\dot{H}^{2}}\delta{\cal G},
\end{equation}
\begin{equation}
	\delta\psi_{2} = \frac{\ell^{2} H^{2}(H^{2}+\dot{H})}{72\dot{H}^{2}}\delta R
	+ \frac{\ell^{2} (2H^{2}+\dot{H})}{144\dot{H}^{3}}\delta {\cal G },
\end{equation}
where we have used the background expressions
\begin{equation}
	\psi_{1} = -1+ \frac{\ell^{2} H^{4}}{6\dot{H}}, \qquad \psi_{2}  = \frac{\ell^{2} H^{2}}{24\dot{H}}.
\label{eq:psiback}
\end{equation}
Then, in terms of the $\Psi$ and $\Phi$ fields we find
\begin{align}
	\psi_{1}(t)+\delta\psi_{1} = -1+\frac{\ell ^2 H^4}{6 \dot{H}}  +  \left(\frac{H^{4}}{\dot{H}^{2}}+\frac{2H^{2}}{\dot{H}} \right)\frac{ \ell^{2}}{9a^{2}}\nabla^{2}\Phi + \frac{\ell^{2}H^{4}}{18 \dot{H}^{2} a^{2}}\nabla^{2}\Psi 
	\nonumber\\
	+\frac{\ell^{2} H^{4}}{6 \dot{H}^{2}}(\ddot{\Phi}+H\dot{\Psi}) - \frac{\ell^{2}H^{3}}{3\dot{H}}(2\dot{\Phi}+ H\Psi),
\end{align}
\begin{align}
	\psi_{2}(t) + \delta\psi_{2} = \frac{\ell^{2} H^{2}}{24\dot{H}} + \left(\frac{H^{2}}{\dot{H}^{2}}+ \frac{1}{\dot{H}} \right)\frac{\ell^{2}}{36 a^{2}}\nabla^{2}\Phi + \frac{\ell^{2} H^{2}}{72\dot{H}a^{2}}\nabla^{2}\Psi + \frac{\ell^{2}H^{3}}{24\dot{H}^{2}}\dot{\Psi}
	\nonumber\\
	- \frac{\ell^{2}H}{12\dot{H}}\dot{\Phi} + \frac{\ell^{2} H^{2} }{24\dot{H}^{2}}\ddot{\Phi}.
\end{align}
and with that the components of Eq. \eqref{eq:mod} are 
\begin{align}
	-6H\left( 1 - \frac{h^{2}}{2} \right)(H\Psi +\dot{\Phi}) + 2\left(  1-\frac{h^{2}}{6} \right)\frac{\nabla^{2}\Phi}{a^{2}}- \frac{2 H h^{2}}{3 \dot{H}}\frac{\nabla^{2}(\dot{\Phi} + H\Psi)}{a^{2}}
	\nonumber\\
	+\frac{h^{2}}{9\dot{H}}\frac{\nabla^{2}\nabla^{2}\Phi}{a^{4}} = 8\pi G_{N}\delta T^{0}_{0},
\label{eq:02}
\end{align}
\begin{align}
	\partial_{i}\Bigg[ 2\left(1-\frac{h^{2}}{2} \right)(\dot{\Phi}+ H\Psi)  - \frac{h^{2}}{9}\left( \frac{H}{\dot{H}} + \frac{\ddot{H}}{\dot{H}^{2}} - \frac{4}{H}\right)\frac{\nabla^{2}\Phi}{a^{2}} + \frac{h^{2}}{9\dot{H}} \frac{\nabla^{2}(\dot{\Phi}+H\Psi)}{a^{2}} \Bigg] = 8\pi G_{N}\delta T^{0}_{i},
\label{eq:03}
\end{align}
\begin{align}
	\Bigg\lbrace-2\left[3H^{2}+2\dot{H}-\frac{h^{2}}{2}(3H^{2}+4\dot{H})   \right]\Psi - 2H\left( 1-\frac{h^{2}}{2} \right)\dot{\Psi}
	\nonumber\\
	- 6H\left[ 1- \frac{h^{2}}{6}\left( \frac{2\dot{H}}{H^{2}} + 3 \right) \right]\dot{\Phi} - 2\left( 1- \frac{h^{2}}{2} \right)\ddot{\Phi}
	\nonumber\\
	+\left[ 1 - \frac{h^{2}}{9}\left( \frac{H^{2}}{2\dot{H}} + \frac{\dot{H}}{H^{2}}+ \frac{\ddot{H}}{H\dot{H}} -  \frac{\ddot{H}^{2}}{\dot{H}^{3}} +  \frac{\overset{...}{H}}{2\dot{H}^{2}} + 1\right)  \right]\frac{\nabla^{2}\Phi}{a^{2}}
	\nonumber\\
	-\left[1- \frac{h^{2}}{9}\left( \frac{5 H\ddot{H}}{2\dot{H}^{2}} -\frac{5 H^{2}}{2\dot{H}} - 6 \right) \right]\frac{\nabla^{2}\Psi}{a^{2}}
	\nonumber\\
	-\frac{h^{2}}{9H^{2}}\frac{\nabla^{2}\nabla^{2}\Phi}{a^{4}} - \frac{h^{2}}{18 \dot{H}}\frac{\nabla^{2}\nabla^{2}\Psi}{a^{4}}
	\nonumber\\
	+\frac{h^{2}}{9}\left( \frac{\ddot{H}}{\dot{H}^{2}} -\frac{3H}{\dot{H}} - \frac{3}{H}  \right)\frac{\nabla^{2}\dot{\Phi}}{a^{2}} - \frac{5 h^{2} H}{18\dot{H}}\frac{\nabla^{2}\dot{\Psi}}{a^{2}} - \frac{5h^{2}}{18\dot{H}}\frac{\nabla^{2}\ddot{\Phi}}{a^{2}}\Bigg\rbrace \delta^{i}_{j}
	\nonumber\\
	- \left[ 1 - \frac{h^{2}}{3}\left( \frac{H^{2}}{2\dot{H}} - \frac{\dot{H}}{H^{2}} + \frac{\ddot{H}}{H\dot{H}} - \frac{\ddot{H}^{2}}{\dot{H}^{3}} + \frac{\overset{...}{H}}{2\dot{H}^{2}}  \right) \right]\frac{\partial^{i}\partial_{j}\Phi}{a^{2}}
	\nonumber\\
	+ \left[  1 - \frac{h^{2}}{6}\left(\frac{H\ddot{H}}{\dot{H}^{2}} - \frac{H^{2}}{\dot{H}} - 2 \right) \right]\frac{\partial^{i}\partial_{j}\Psi}{a^{2}}
	\nonumber\\
	+\frac{h^{2}}{9H^{2}}\frac{\partial^{i}\partial_{j}\nabla^{2}\Phi}{a^{4}} + \frac{h^{2}}{18 \dot{H}}\frac{\partial^{i}\partial_{j}\nabla^{2}\Psi}{a^{4}}
	\nonumber\\
	+\frac{h^{2}}{3}\left( \frac{H}{\dot{H}} - \frac{1}{H} \right)\frac{\partial^{i}\partial_{j}\dot{\Phi}}{a^{2}} +\frac{h^{2}H}{6\dot{H}}\frac{\partial^{i}\partial_{j}\dot{\Psi}}{a^{2}} = 8\pi G_{N}\delta T^{i}_{j}.
\label{eq:4}
\end{align}

In the Fourier space Eqs.\ \eqref{eq:02} and \eqref{eq:03} can be written respectively  in the form
\begin{align}
	\left( -6 +3 h^{2}  + \frac{2h^{2}k^2}{3a^2\dot{H}}  \right) H(\dot{\Phi} + H \Psi) + \left(  -2 + \frac{h^{2}}{3}   + \frac{h^{2}k^2}{9a^2\dot{H}}   \right)\ \frac{k^2}{a^2}\Phi  =8\pi G_{N}\delta T^{0}_{0} ,
\label{eq:7}
\end{align}
\begin{align}
	\left[\left( 2 -h^2   - \frac{h^{2}k^2}{9a^2\dot{H}}  \right)(\dot{\Phi} + H \Psi) + \frac{h^{2}}{9}\left( \frac{H}{\dot{H}}+\frac{\ddot{H}}{\dot{H}^{2}} - \frac{4}{H}  \right)\frac{k^2}{a^2} \right](ik_{i})= 8\pi G_{N}\delta T^{0}_{i} .
	\label{eq:8}
\end{align}

\section{Tensor perturbations}
\label{app-tensor}
This appendix deals with tensor mode perturbations and has a structure similar to the one in appendix \ref{app-scalar}.
 For tensor modes, the perturbed line element is given by
\begin{equation}
ds^{2} = dt^{2} - a^{2}(\delta_{ij} + h_{ij})dx^{i}dx^{j},
\label{eq:139}
\end{equation}
where $h_{ij}$ is traceless and transverse tensor.
In the next section we will provide the expressions for the geometric quantities,
and in Sec. \ref{tensorperteq}  we derive the the perturbed field equations.

\subsection{Perturbed geometric quantities}
\label{perturbed}

The elements of the metric tensor and its inverse can be
straightforwardly obtained from \eqref{eq:139}.
Plugging them into the metric connection one gets the following expressions
for the non-null components of Christoffel symbols:
\begin{eqnarray}
\Gamma^{0}_{ij}   = a^{2}H \delta_{ij} +a^{2}H h_{ij}+\frac{a^{2}}{2}\dot{h}_{ij},
\\
\Gamma^{i}_{0j}   = H\delta^{i}_{j}+ \frac{\delta^{ik}}{2}\dot{h}_{kj},
 \\
\Gamma^{i}_{jk}   = \frac{\delta ^{in}}{2}(\partial_{j}h_{kn}+\partial_{k}h_{nj} - \partial_{n}h_{jk}).
\end{eqnarray}
The linear parts of the covariant, contravariant and mixed components of interest of
the Ricci tensor are 
\begin{equation}
\delta R_{ij} = a^{2}(3H^{2}+\dot{H})h_{ij} + \frac{3}{2}a^{2}H\dot{h}_{ij}
+ \frac{a^{2}}{2}\ddot{h}_{ij} - \frac{1}{2}\nabla^{2}h_{ij},
\end{equation}
\begin{equation}
\delta R^{ij} = \frac{\delta^{ik}\delta^{jl}}{a^{2}}\Bigg(- (3H^{2} +\dot{H})h_{kl}
+ \frac{3}{2}H \dot{h}_{kl} +\frac{1}{2}\ddot{h}_{kl} - \frac{1}{2a^{2}}\nabla^{2}h_{kl}\Bigg),
\end{equation}
\begin{equation}
\delta R^{i}_{j}= \frac{\delta^{ik}}{2}\left( \frac{1}{a^{2}}\nabla^{2}h_{kj}
- 3 H \dot{h}_{kj} -\ddot{h}_{kj}    \right),
\end{equation}
and the linear parts of the Riemann tensor are
\begin{equation}
\delta R_{0i0j} =  a^{2} \left( (H^{2}+\dot{H})h_{ij}+ H\dot{h}_{ij} +\frac{1}{2}\ddot{h}_{ij}  \right),
\end{equation}
\begin{equation}
\delta R^{i}_{0j0} =  -\delta^{ik}\left(H\dot{h}_{kj} +\frac{1}{2}\ddot{h}_{kj}\right),
\end{equation}
\begin{eqnarray}
\delta R^{i}_{jkl} =    \frac{\delta^{in}}{2} (\partial_{k}\partial_{j}h_{ln}
- \partial_{k}\partial_{n}h_{jl} - \partial_{l}\partial_{j}h_{kn} +\partial_{l}\partial_{n}h_{jk} )
+ \frac{\delta^{in}\dot{h}_{nk}}{2}a^{2}H\delta_{lj}\nonumber
 \\
    +H\delta^{i}_{k}\left( a^{2}H h_{lj}+ a^{2}\frac{\dot{h}_{lj}}{2} \right)
 - \frac{\delta^{in}\dot{h}_{nl}}{2}a^{2}H\delta_{kj} - H\delta^{i}_{l}\left( a^{2}Hh_{kj}+ a^{2}\frac{\dot{h}_{kj}}{2} \right),
\end{eqnarray}
\begin{equation}
\delta R_{mjkl} = - a^{4}H  (h_{mk}\delta_{lj} - h_{ml}\delta _{kj}) - a^{2}\delta_{im}\delta R^{i}_{jkl},
\end{equation}
where we have used the background expressions for  the geometric quantities as in Sec.\ \ref{sec:pgq}.

\subsection{Perturbed field equations}\label{tensorperteq}

From the equations derived in Sec.\ \ref{perturbed} one can easily conclude that
tensor modes do not induce fluctuations in the Ricci scalar, Gauss-Bonnet
invariant and $\Box \psi_{b}$.  Therefore, the space-space component of \eqref{eq:mod} are
\begin{eqnarray}
  -\psi_{1}\delta R^{i}_{j}  + \delta(g^{i\sigma}\nabla_{\sigma}\nabla_{j}\psi_{1})
+ 4(\Box\psi_{2})\delta R^{i}_{j} + 2 \delta(g^{i\sigma}\nabla_{\sigma}\nabla_{j}\psi_{2})\overset{}{R}
- 4\delta(g^{i\sigma}\nabla_{\rho}\nabla_{\sigma}\psi_{2})\overset{}{R^{\rho}_{j}} \nonumber
\\
  -4(\overset{}{g^{i\sigma}}\nabla_{\rho}\nabla_{\sigma}\psi_{2})\delta R^{\rho}_{j}
- 4\delta(\nabla_{\rho}\nabla_{j}\psi_{2})\overset{}{R^{\rho i}}
- 4(\nabla_{\rho}\nabla_{j}\psi_{2})\delta R^{\rho i}
- 4\delta(\nabla^{\rho}\nabla^{\sigma}\psi_{2})\overset{}{R^{i}_{\rho j \sigma}} \nonumber
\\
  + 4\delta^{i}_{j} \big[  \delta(\nabla_{\rho}\nabla_{\sigma}\psi_{2})\overset{}{R^{\rho\sigma}}
+ (\nabla_{\rho}\nabla_{\sigma}\psi_{2})\delta R^{\rho\sigma}  \big]
 - 4(\nabla^{\rho}\nabla^{\sigma}\psi_{2})\delta R^{i}_{\rho j \sigma} = 8\pi G \delta T^{i}_{j}.
 \label{eq:150}
\end{eqnarray}
 For tensor modes $\delta (\nabla_{\alpha}\nabla_{\beta} \psi_{b}) = -\delta \Gamma^{0}_{\alpha\beta}\dot{\psi}_{b}$.
Using this, the terms in the above equation can be expressed as
\begin{equation}
\delta( g^{i\sigma}\nabla_{\sigma}\nabla_{j}\psi_{1}) =\frac{ \dot{\psi}_{1}}{2}\delta^{ik}\dot{h}_{kj},
\end{equation}
\begin{equation}
\delta (g^{i\sigma}\nabla_{\rho}\nabla_{\sigma}\psi_{2})R^{\rho}_{j} =
- \frac{\dot{\psi}_{2}}{2}(3H^{2}+\dot{H})\delta^{ik}\dot{h}_{kj},
\end{equation}
\begin{equation}
(g^{i\sigma}\nabla_{\rho}\nabla_{\sigma}\psi_{2})\delta R^{\rho}_{j} =  H\dot{\psi}_{2}\delta R^{i}_{j},
\end{equation}
\begin{equation}
\delta(\nabla_{\rho}\nabla_{j}\psi_{2})R^{\rho i} =  - \dot{\psi}_{2} (3H^{2}+\dot{H})\delta^{ik}\left(H h_{kj}
+ \frac{1}{2}\dot{h}_{kj}\right),
\end{equation}
\begin{equation}
(\nabla_{\rho}\nabla_{j}\psi_{2})\delta R^{\rho i} = -a^{2}H\dot{\psi}_{2}\delta_{kj}\delta R^{ki},
\end{equation}
\begin{equation}
\delta(\nabla^{\rho}\nabla^{\sigma}\psi_{2})R^{i}_{\rho j \sigma}  =
- H^{2}\dot{\psi}_{2}\delta^{ik}\left(  H h_{kj} -\frac{\dot{h}_{kj}}{2}   \right),
\end{equation}
\begin{equation}
(\nabla^{\rho}\nabla^{\sigma}\psi_{2})\delta R^{i}_{\rho j \sigma}
= \frac{H\dot{\psi}_{2}}{2 a^{2}}\delta^{ik}\nabla^{2}h_{kj}
- H\left(\ddot{\psi}_{2}+\frac{H\dot{\psi}_{2}}{2}\right)\delta^{ik}\dot{h}_{kj}
+\dot{\psi}_{2}H^{3}\delta^{ik}h_{kj} - \frac{\ddot{\psi}_{2}}{2}\delta^{ik}\ddot{h}_{kj}.
\end{equation}
Then, using these expressions in Eq.\ \eqref{eq:150} we obtain
\begin{eqnarray}
\Bigg[  \frac{1}{2}\left( \psi_{1} - 4H\dot{\psi}_{2} \right)\ddot{h}_{kj} + \frac{1}{2}\left( 3H\psi_{1}
+ \dot{\psi}_{1} - 4H\ddot{\psi}_{2} - 4(3H^{2}+\dot{H})\dot{\psi}_{2} \right)\dot{h}_{kj}
\nonumber\\
  - 8H^{3}\dot{\psi}_{2}h_{kj} - \frac{1}{2}(\psi_{1}-4\ddot{\psi}_{2})\frac{\nabla^{2}h_{kj}}{a^{2}}\Bigg]\delta^{ik}
= 8\pi G \delta T^{i}_{j}.
\end{eqnarray}
By substituting the background expressions \eqref{eq:psiback} for the auxiliary fields  we obtain
\begin{eqnarray}
\left[
- \frac{1}{2} \left( 1- \frac{h^{2}}{6}\left( \frac{H^{2}}{\dot{H}} + \frac{H\ddot{H}}{\dot{H}^{2}}  -2 \right)
\right)\ddot{h}_{kj} \right.
\nonumber\\
-\frac{3}{2}H\left( 1- \frac{h^{2}}{6}\left( \frac{H^{2}}{\dot{H}} - \frac{4\dot{H}}{3 H^{2}} +
\frac{2 H\ddot{H}}{3\dot{H}^{2}} +\frac{\ddot{H}}{H\dot{H}} - \frac{2\ddot{H}^{2}}{3\dot{H}^{3}}+
\frac{\overset{...}{H}}{3\dot{H}^{2}}  - \frac{2}{3}\right)   \right)\dot{h}_{kj}
\nonumber\\
-H^{2}\frac{h^{2}}{3}\left( 2 -\frac{H\ddot{H}}{\dot{H}^{2}}  \right)h_{kj}
\nonumber\\
\left. +\frac{1}{2}\left( 1 - \frac{h^{2}}{6} \left( \frac{H^{2}}{\dot{H}}
 - \frac{2 \dot{H}}{H^{2}} +\frac{2\ddot{H}}{H\dot{H}} -
\frac{2\ddot{H}^{2}}{\dot{H}^{3}} +\frac{\overset{...}{H}}{\dot{H}^{2}}    \right)
\right)\frac{\nabla^{2}h_{kj}}{a^{2}}\right] \delta^{ik}= 8\pi G \delta T^{i}_{j}.
\label{eq5200}
\end{eqnarray}
In the limit $h^2\equiv\ell^2 H^2\rightarrow 0$
this equation takes the usual general relativity form.

\section{Ghost instabilities in $F(R,\mathcal{G})$ theory}
\label{ghost}
A general action
\begin{equation}
	S= \frac{1}{16\pi G_{\rm N}}\int d^{4}x\sqrt{-g}F(R,P,Q)+\int d^{4}x\sqrt{-g}\mathcal{L}_{\rm matt},
	\label{eq1}
\end{equation}
where 
\begin{equation}
	P = R^{\mu\nu} R_{\mu\nu} , \quad
		Q = R^{\mu\nu\rho\sigma} R_{\mu\nu\rho\sigma}
	\label{eq2}
\end{equation}
can be expanded around a background 
up to second order in the fluctuations.
It has been shown \cite{hindawi,chiba} that such an expansion  will be identical to that obtained from 
\begin{equation}
	S= \frac{1}{16\pi G_{\rm N}}\int d^{4}x\sqrt{-g}\left( -2 \Lambda + a R + \frac{b}{2} R^2
	-\frac{c}{6} C^2 \right)+\int d^{4}x\sqrt{-g}\mathcal{L}_{\rm matt} ,
	\label{eq5}
\end{equation}
where 
\begin{equation}
	C^2\equiv  C^{\mu\nu\rho\sigma}C_{\mu\nu\rho\sigma} =
	Q-2P+\frac13 R^2
	\label{eq6}
\end{equation}
is the Weyl tensor squared and the coefficients
$\Lambda$, $ a $, $b$, and $c$  
depend on the background.

The gravity theory with the four-derivative terms $R^2$ and $C^2$  was
extensively studied \cite{stelle,hindawi,chiba,bogdanos,martino,comelli,navarro} and it was found that there appear new degrees of
freedom in addition to the spin-2 massless graviton: a massive spin-0 field (with mass $m_0^2\propto 1/b$ corresponding to the $R^2$ term  and a massive spin-2 field (with mass $m_2^2\propto 1/c$) corresponding to the
Weyl squared term. Moreover, the massive spin-2 field is known to have a wrong sign of
the kinetic term and thus has negative energy: a ghost field.

Consider first the
Minkowski background. 
If we linearize gravity using $g_{\mu\nu}=\eta_{\mu\nu}+h_{\mu\nu}$,
the traceless part of the metric perturbation $\bar{h}_{\mu\nu}$  in momentum space will satisfy
\begin{equation}
	\left(k^2-\frac{k^4}{m_2^2}\right)\bar{h}_{\mu\nu}=0 ,
	\label{eq11}
\end{equation}
where
\begin{equation} 
	m_2^2= - \frac{F_{R}}{F_{P}+4F_{Q}},
	\label{eq02}
\end{equation}
with  $F_R=\partial F/\partial R$ etc., and it is understood that these derivatives are evaluated on the background.
Note that the propagator for $\bar{h}_{\mu\nu}$ can be written as \cite{bogdanos} 
\begin{equation}
	G(k) \propto \frac{1}{k^2-k^4/m_2^2}
	=\frac{1}{k^2}-\frac{1}{k^2-m_2^2} .
	\label{eq13}
\end{equation}
The first term on the right-hand side corresponds to the massless graviton whereas the second term corresponds to the massive spin-2 field with mass $m_2$.
The second term has the opposite sign, which indicates the presence of a ghost.
Hence, ghost terms may occur  for a general $F(R, P,Q)$ theory and is parameterized by  the $k^4$  mode.
However, for a theory of the form  $F=F(R,4P-Q)$  we have $F_P+4F_Q=0$
and  equation (\ref{eq11})  simplifies to 
$k^2\bar{h}_{\mu\nu}=0$. Clearly
in this case  the mentioned spin-2 ghost is absent.

Next, assume that the background is a constant curvature
maximally symmetric spacetime.
In this case we have $R_{\mu\nu\rho\sigma}=(R/12) (g_{\mu\rho}g_{\nu\sigma}-g_{\mu\sigma}g_{\nu\rho})$ so
$P=R^2/4$ and $Q=R^2/6$.
The propagator of the graviton has the same structure as in  (\ref{eq13}) with  the ghost mass 
 $m_2^2\propto -({F_{P}+4F_{Q}})^{-1}$,
 similar to (\ref{eq02}).
As before, the presence of the Weyl  term implies the presence of a ghost field.
Again, for the theory where $F=F(R,4P-Q)$ we have $c=0$,  and the $C^2$ term in (\ref{eq5}) is absent. We are left with an effective $R^2$ theory in which there are no ghosts. This applies also to
a general  $F(R,\mathcal{G})$ theory 
including our model 
\begin{align}
	F(R,{\cal G}) = -R + f(J),
	\label{eq700}
\end{align}
where $J$ is defined in Eq.\ (\ref{eq1003}).

There is still some ambiguity concerning a possible 
presence of instabilities in the scalar sector. 
De Felice and Suyama  \cite{felice} (hereafter DFS) argue that
an instability 
can arise in vacuum for the scalar modes of the cosmological perturbations if the background is not
de Sitter.  The cause of instability is a term proportional to $k^4$,
which, apart from a few exceptions (examples are provided in their Table 1), appears in their  master equation (see below).
In contrast, Navarro and Van Acoleyen \cite{navarro} find  no such $k^4$ instability in the scalar sector 
for a general $F(R, 4P-Q)$. They derive a propagator for the scalar field but their derivation in appendix 
is restricted to de Sitter space although they argue that in a generic FLRW background the results 
will be qualitatively similar.
This casts a certain doubt on the validity of their result.

The analysis of DFS is based on the following equation 
\begin{equation}
	\frac{1}{a^3 Q} \partial_t (a^3 Q \dot {\hat \Phi}) + B_1 \frac{k^2}{a^2} \hat \Phi + B_2 \frac{k^4}{a^4} \hat \Phi = 0 ,  
\label{masterDFS}
\end{equation}
dubbed  ``master equation'' (Eq.\ (59) in \cite{felice}).
The quantities $Q$, $B_1$, and $B_2$ are time dependent background functions and $\hat \Phi$ is a gauge invariant scalar perturbation in the spatial slices of constant time.
The quantity $\hat \Phi$ is one of the three gauge invariant  perturbations introduced in Ref.\ \cite{felice} to describe general scalar perturbations for generic $F(R,{\cal G})$ theories. 
In our notation
\begin{align}
	\hat{\Phi}   = \Phi + \frac{H(\delta\psi_{1} - 4H^{2}\delta\psi_{2})}{\dot{\psi_{1}}-4H^{2}\dot{\psi_{2}}} ,
\label{eq:invvar}
\end{align}
where $\psi_1$ and $\psi_2$ are auxiliary fields defined in Sect.\ \ref{scalar-tensor}.
By manipulating Eqs.\ (\ref{eq:130}),  (\ref{eq:131}), 
and Eq.\ (\ref{eq:132}) 
it is possible to eliminate two other scalars in favor of $\hat \Phi$, thus yielding the master equations 
(\ref{masterDFS}).
 If a solution  to (\ref{masterDFS})
 exhibits an instability (e.g., exponential growth of $\hat \Phi$), this implies that the scalar perturbations of the $F(R, {\cal G})$ model are plagued by an instability. 

 The problematic term in Eq.\ \eqref{masterDFS} is the one that contains the $k^4$ term. 
 	For models where this term vanishes, i.e., when $B_2=0$,
 	we have a standard wave equation. For a general $F(R, {\cal G})$ model, $B_2=0$ if the background is de Sitter.
 	Besides, 
 	 coefficient $B_2$ is identically  zero for some particular cases of $F(R, {\cal G})$, e.g., for $F = f(R) + \mbox{const.}\, {\cal G}$ or $F = f(R + \mbox{const.}\, {\cal G})$ \cite{felice}.
 Note that the models of inflation  considered in Refs.\ \cite{delaurentis,odintsov4} belong to the class of models in which 
 $B_2$ does not  vanish identically.

For short wavelengths,  equation \eqref{masterDFS} is solved  using the WKB approximation and the properties of the solution are discussed in detail \cite{felice}. In a nutshell, if $B_{2}$ is negative $\hat{\Phi}$ grows exponentially with time, implying that the perturbations in FLRW universe is unstable on small scales. These instabilities grow to the point where linear perturbations are not valid.
If $B_{2}$ is positive, the perturbations propagate with group velocity 
\begin{align}
	v_{g}(k)\approx 2\sqrt{B_{2}}\frac{k}{a}.
\end{align}
The group velocity
exceeds the speed of light  for modes above a critical $k_{\rm cr}\approx a/(2\sqrt{B_2})$ so the short wavelength  propagation modes will be superluminal for $k>k_{\rm cr}$.  
 However,  a superluminal behavior of the cosmological perturbations is not necessarily  unphysical \cite{Babichev:2007dw}.
Note that a superluminal behavior is absent in a theory in which the critical $k_{\rm cr}$ exceeds the cutoff of the theory $\Lambda_{\mathrm{cutoff}}$.

Finally, we  investigate the coefficients $B_1$ and $B_2$ 
for  models of the type (\ref{eq700}).
We have computed these factors for a general $F(R,{\cal G})$ model in Newtonian gauge and our results agree with those obtained in Ref.\ \cite{felice} where the
precise expressions can be found. 
The point essential for our analysis is that both $B_1$ and $B_2$ can be expressed as fractions  the denominators of which contain a factor 
\begin{align}
	D=	2\dot{H}(2H^{2}\dot{\psi}_{1} +\dot{H}\dot{\psi}_{1}-48H\dot{H}^3\psi_{1,{\cal G}}  - 8H^{4}\dot{\psi}_{2}) +H\ddot{H}(\dot{\psi}_{1}-4H^{2}\dot{\psi}_{2}). \label{eq:null}
\end{align}
It may be explicitly verified that
this factor 
 vanishes identically  
for  models  of the type \eqref{eq700}.
To see this, consider 
the background quantities 
\begin{align}
	\psi_{1} = \frac{\partial F}{\partial R} = -1 + \frac{\partial f}{\partial J}\frac{\partial J}{\partial R}, \qquad \psi_{2} = \frac{\partial F}{\partial {\cal G}} =\frac{\partial f}{\partial J}\frac{\partial J}{\partial {\cal G}}.
\end{align}
Inserting  the 
time derivatives of $\psi_1$  and $\psi_2$ and the derivative of $\psi_1$ with respect to $\mathcal{G}$ into \eqref{eq:null} one finds that the factor $D$  is identically zero. 
This implies
that the coefficients $B_{1}$ and $B_{2}$
 are ill defined and the master equation \eqref{masterDFS} cannot serve as a check for ghost instabilities in a model of the type (\ref{eq700}). In other words,  the DFS analysis is inconclusive with regard to the presence of instabilities in models of the type (\ref{eq700}).

\end{document}